\documentclass{aa}  
\usepackage{subcaption} 
\usepackage{xcolor, colortbl} 
\usepackage[normalem]{ulem}   

 \usepackage{multirow}
 \usepackage{booktabs}
\usepackage{graphicx}
\usepackage{txfonts}
\usepackage{hyperref}
    \hypersetup{colorlinks=true, linkcolor=blue, citecolor=blue, urlcolor=blue}

\usepackage{siunitx}

\begin{document}

   \title{\texttt{TILARA}: Template-Independent Line-by-line Algorithm for Radial velocity Analysis}

   \subtitle{I. Description of the code and application on a Sun-like star}

   \author{C. San Nicolas Martinez\inst{1,2}
           \and N. C. Santos\inst{1,2}
          \and V. Adibekyan\inst{1,2}
          \and K. Al Moulla\inst{1}\thanks{SNSF Postdoctoral Fellow}
          \and A. M. Silva\inst{1,2}
          \and S. G. Sousa\inst{1,2}}

   \institute{Instituto de Astrof\'isica e Ci\^encias do Espa\c{c}o, Universidade do Porto, CAUP, Rua das Estrelas, 4150-762 Porto, Portugal
             \and Departamento de F\'isica e Astronomia, Faculdade de Ci\^encias, Universidade do Porto, Rua do Campo Alegre, 4169-007 Porto, Portugal
             }

   \date{Received November 13, 2025; accepted March 4, 2026}

 
  \abstract
   {Precise radial velocities (RVs) are commonly derived through cross-correlation or template-matching methods, both of which rely on a reference spectrum that can introduce biases when the data are variable, contaminated, or sparsely sampled. Line-by-line methods offer an alternative way to compute RVs but generally still rely on template creation and therefore share its inherent limitations.}
   {We introduce \texttt{TILARA}, a template-independent, line-by-line RV extraction code designed to allow us to derive line-by-line RVs and to operate effectively even when spectral template construction is not recommended. While originally motivated by future PoET disk-resolved solar observations, \texttt{TILARA} has been built with the flexibility to work with different stellar spectral types and instruments.}
   {A curated list of individual absorption lines is used as a reference to automatically measure line centers with via Gaussian fitting with \texttt{ARES}. Then, using the reference lines list, and the lines measured with \texttt{ARES} on the spectra of the target star, \texttt{TILARA} computes the RVs and applies configurable outlier rejection through sigma-clipping or down-weighting methods.}
   {We tested different configurations of the code, RV uncertainty estimation methods, and line selection criteria. The code was applied to 520 ESPRESSO observations of the Sun-like star HD~102365 to evaluate its performance.
\texttt{TILARA} was then tested against other RV extraction methods. Both in its sigma-clipping and its down-weighting mode, \texttt{TILARA} provided resulting RV time-series with similar standard deviation and error bars as the ones derived using existing methods that follow different approaches.}
   {Our results demonstrate that \texttt{TILARA} can deliver precise, template-free, RV measurements and effectively reduce the impact of spectral outliers. Its flexibility makes it well suited for both current ultra-precise spectrographs and future applications to spatially resolved solar observations. Additionally, the ability to easily modify the linelist allows \texttt{TILARA} to be adapted for exploring how various spectral features or physical phenomena affect RV measurements.}


   \maketitle
%

\section{Introduction}
   
High-resolution Doppler spectroscopy remains a cornerstone in exoplanet detection, particularly when combined with transit photometry \citep{Mayor2014,Santos2020,faria2022}, as together they provide key information on a planet’s mass, radius, and density. Instruments such as ESPRESSO, capable of reaching radial velocity (RV) precisions of 10~cm\,s$^{-1}$ -- comparable to the signal induced by an Earth-like planet orbiting a Sun-like star \citep{cegla2019} -- represent major technological advances in the field \citep{Pepe2021}.  

Despite this progress, stellar noise, originating from magnetic activity, granulation, oscillations, and other processes, remains a significant obstacle \citep{Schrijver2008,lovis2011,Santos2014, Borgniet2015,Gilbertson2020,Camacho2022, Burt2025}. A major breakthrough expected in the near future is the first set of observations from the Paranal solar ESPRESSO Telescope \citep[PoET;][]{Santos2025}, an instrument capable of performing disk-resolved observations of the Sun. This capability will be key for understanding and disentangling stellar variability from planetary signals on a RV dimension.  

Several methods are currently used to compute RVs. One common approach is the cross-correlation function (CCF) \citep{Baranne1996,griffin2000,Pepe2002}. Another common method is template matching, which can either provide a single RV measurement for the full spectrum (or for full spectral orders) \citep[e.g.,][]{Angladaescude2012,Zechmeister2018,Silva2022}, or it can be used to measure the RVs of individual spectral lines \citep[e.g.,][] {Dumusque2018,Cretignier2020,Almoulla2022,Artigau2022}. Line-by-line RV measurements have been shown to be outlier resistant and useful to probe stellar variability that affects certain subset of spectral lines. However, since most commonly employed line-by-line methods use template matching to obtain the RVs, they are by design restricted by the associated limitations. Template matching can struggle when template creation is not feasible, for instance:

   \begin{figure*}
   \centering
   \includegraphics[width=1\textwidth]{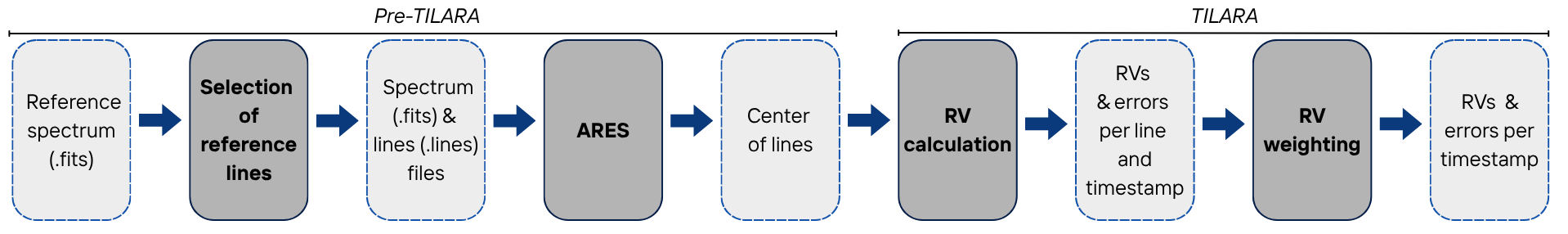}
   \caption{Flowchart of  \texttt{TILARA} steps. Dark grey bubbles indicate computational steps, and light grey bubbles show intermediate inputs/outputs. The first two steps correspond to the work done with \texttt{linesearcher} and \texttt{ARES} before using \texttt{TILARA}, while the last two steps correspond to the work done by \texttt{TILARA}.} 
              \label{Fluxogram}%
    \end{figure*}
  
\begin{itemize}
    \item When there are not enough observations to build a sufficiently high signal-to-noise ratio (S/N) template, making the measured RVs noisier and requiring the use of a template from a similar star observed with the same instrument.
    \item When the template is built from observations taken over a short time span, which can lead to systematic trends in the measured RVs associated with interpolation errors and/or telluric contamination \citep{Doshi2025,Silva2025}.
    \item When studying high spatial-resolution surface phenomena, such as those that will be observed with the upcoming PoET telescope, the spectral variations are too strongly in time for a stable reference template to exist, breaking the fundamental assumption of template-matching and therefore current line-by-line methods, which require a fixed spectral shape to measure Doppler shifts.

\end{itemize}

Furthermore, as long as spectral synthesis is unable to reproduce stellar spectra to the accuracy required to measure centimeter-per-second velocity variations, templates are necessarily built from observations, which means the templates themselves can currently never be entirely free from shifts and asymmetries introduced by stellar variability.

To overcome these limitations, a broad class of template-independent RV estimators has been developed that does not rely on a fixed reference spectrum. These methods can operate at a global spectrum level \citep{zucker2006}, or on a line-by-line basics. Operating at the level of individual spectral features, these methods give rise to so-called line-by-line (LBL) RV techniques. These include forward-modeling and data-driven methods that analyze spectral chunks or line-shape variability to disentangle stellar activity from Doppler shifts, such as those developed within the EXPRES Stellar Signals Project \citep[e.g.,][]{Zhao2022}. Other studies directly fit individual spectral features: \citet{Liebing2021} model line-core shifts to probe convective blueshift, \citet{Lafarga2023} perform Gaussian fits to thousands of lines to identify subsets with reduced activity sensitivity, and \citet{GomesdaSilva2025} adopt parametric line-profile fitting in the near-infrared.

Despite their promise, these template-independent line-by-line approaches still face significant limitations. No method has yet demonstrated a consistent reduction of RV scatter to sub-meter-per-second precision across different stars, and results often vary substantially between techniques \citep{Zhao2022}. Several approaches are primarily sensitive to line-shape variability and remain unable to disentangle pure translational Doppler shifts, such as those induced by stellar oscillations. Others are strongly constrained by instrumental resolution, S/N, and stellar properties, and can reach intrinsic noise floors set by granulation variability \citep{Liebing2021}. In addition, activity sensitivity at the line level is not universal across stars, limiting the applicability of fixed \textit{clean} line lists and often resulting in RV uncertainties larger than those obtained with traditional methods.

Building on these previous works, and to help mitigate the limitations of the line-by-line method while keeping all its strengths, we introduce \texttt{TILARA} (Template-Independent Line-by-line Algorithm for Radial velocity Analysis), which works in combination with \texttt{ARES} \citep{Sousa2007,Sousa2015} to obtain line-by-line RV measurements based on Gaussian fitting. By avoiding reliance on a pre-constructed template, \texttt{TILARA} addresses the limitations of both template matching and existing line-by-line methods, enabling robust RV extraction when a stable reference spectrum cannot be defined. In this context, \textit{template-independence} refers to the extraction process itself: unlike traditional methods that shift a flux profile to minimize residuals, \texttt{TILARA} treats each absorption feature as an independent entity. While a list of reference wavelengths is required to identify the features of interest, the algorithm is agnostic to the source of this list -- be it empirical observations or synthetic catalogs -- and does not rely on the flux shape of a reference template during the RV calculation. An additional advantage of \texttt{TILARA} is that it measures true individual spectral lines: while other line-by-line codes define "lines" as segments between consecutive local maxima, \texttt{ARES} can fit multiple Gaussian components within blended spectral features when they are separable in the observed spectrum, allowing \texttt{TILARA} to study individual absorption components. This also gives the user full control over which subsets of lines are used, enabling tailored RV extractions based on line depth, formation height, activity sensitivity, or any other physical criterion.

This paper is organized as follows. In Sect.~\ref{sec:TILARA} we present the \texttt{TILARA} code, describing its main modules and the methodology adopted for the extraction of line-by-line RVs. Sect.~\ref{HD~102365} details the application of \texttt{TILARA} to ESPRESSO observations of the Sun-like star HD~102365, including a comparison with other state-of-the-art RV extraction pipelines. In Sect.~\ref{sec:conclusions} we summarize the main results and discuss the advantages and future prospects of the \texttt{TILARA} approach, particularly in the context of upcoming disk-resolved solar observations. Additional technical details, tests, and complementary analyses are provided in the Appendices.




\section{The \texttt{TILARA} code} \label{sec:TILARA}


%
\texttt{TILARA} is designed to extract precise RV measurements from spectroscopic observations. The combination of \texttt{TILARA} and \texttt{ARES} workflow operates through four main steps (Fig.~\ref{Fluxogram}):  
\begin{enumerate}
    \item \textbf{Selection of reference lines:} Uses \texttt{linesearcher}\footnote{\url{https://github.com/sousasag/linesearcher}} and \texttt{ARES}\footnote{\url{https://github.com/sousasag/ARES}} to build the reference line list (\texttt{.lines}) from reference spectrum (in this paper, solar spectra).
    \item \textbf{\texttt{ARES}:} \texttt{ARES} takes the reference line list together with the observations of the target star (in this paper the ESPRESSO \textbf{Data Reduction Software} (DRS) files \citep{Pepe2021}\footnote{The ESPRESSO DRS is the pipeline that converts raw instrument FITS frames into calibrated data products.} to determine the center of the spectral lines -- for the subsequent RV calculations -- and measure their parameters across all observations.
    \item \textbf{RV calculation:} Computes line-by-line RVs and their associated error for each timestamp.
    \item \textbf{RV weighting:} Derives epoch-wise RVs by calculating a weighted mean of the individual line-by-line RVs.
\end{enumerate}

\subsection{Step 1: Selection of reference lines} \label{Selection_of_lines}

The first step of the process is to obtain a curated line list that provides the wavelengths that are later used to compute the RVs. In this paper we obtain it from a reference spectrum built from 5 solar observations (see in Section~\ref{Solarlines}) but the line list can also be user provided. While in this paper we use the proposed method -- combining \texttt{linesearcher} and \texttt{ARES} -- to generate the reference line list, users may provide their own list obtained by other means, as long as the input for \texttt{ARES} in \textit{Step~2} is a file with two columns, wavelength in the first and line depth in the second.

\texttt{ARES}\footnote{The \texttt{ARES} output was adapted for \texttt{TILARA} to provide higher numerical precision in the output parameters.} measures the central wavelengths of the absorption lines used in \texttt{TILARA}  to compute the RVs. For this, \texttt{ARES} requires one-dimensional (S1D) spectra and the input of a premiliminary list of spectral lines suitable for the target star. The first stage of the workflow therefore consists in building this line list, which we achieve by combining \texttt{linesearcher} and \texttt{ARES} (see Appendixes ~\ref{app:line_selection} and ~\ref{app:ARES} for additional details on the codes performance).

\texttt{Linesearcher} -- a code that finds the lines positions in a spectrum  -- is first used to identify candidate spectral features and provide initial estimates of their central wavelengths and depths. In \texttt{linesearcher}, all spectra are locally normalized prior to line fitting, such that the continuum level is close to unity, and line depths are therefore defined relative to a normalized flux scale. These lines are then re-analyzed with \texttt{ARES}, a code that fits Gaussian profiles to individual features and outputs their equivalent widths (EWs), full width at half maximum (FWHM), depths, and associated error \citep{Sousa2007}. 
In its updated version, \texttt{ARES} also performs automatic RV correction and improved continuum normalization, increasing the robustness of the measurements \citep{Sousa2015}.

The result of this step is a cleaned and homogenized list of absorption lines \textbf{--} that in this paper are consistently detected across the solar spectra \textbf{--} and will serve as the reference set for the line-by-line RV extraction. The full details of the preprocessing, parameter choices, and selection filters applied to the line list are provided in Appendices~\ref{app:line_selection} and~\ref{app:ARES}.

\subsubsection{Reference lines for Sun-like stars} \label{Solarlines}

 \begin{figure}
   \centering
   \includegraphics[width=0.5\textwidth]{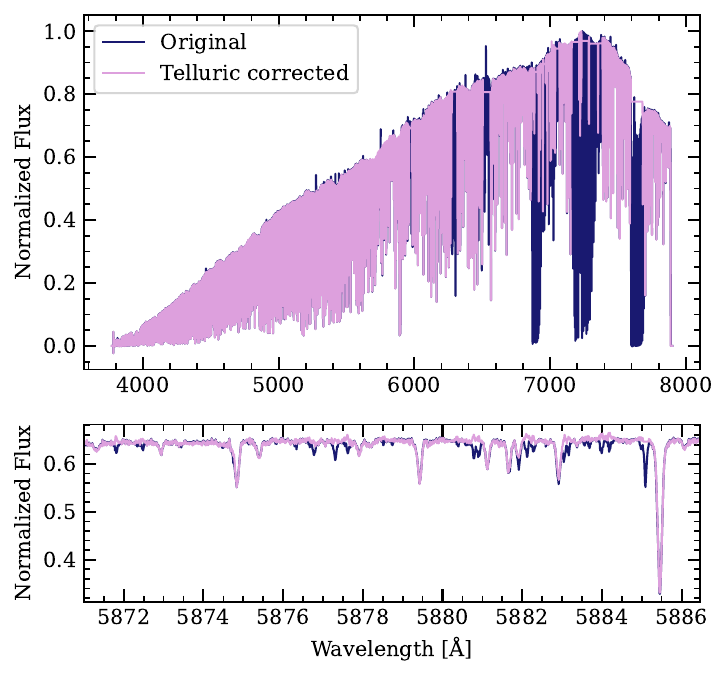}
   \caption{Example of the telluric correction done with \texttt{Molecfit} applied to the solar ESPRESSO spectra. The dark violet spectrum shows the original observation, while the lilac spectrum displays the same observation after the removal of telluric H$_2$O, O$_3$ and O$_2$ features.
   \textit{Top panel:} Full spectral range.  \textit{Bottom panel:} Zoomed-in wavelength range between 5871 and 5886 $\AA$.}
              \label{telluric}%
    \end{figure}
 \begin{figure*}
   \centering
   \includegraphics[width=1.0\textwidth]{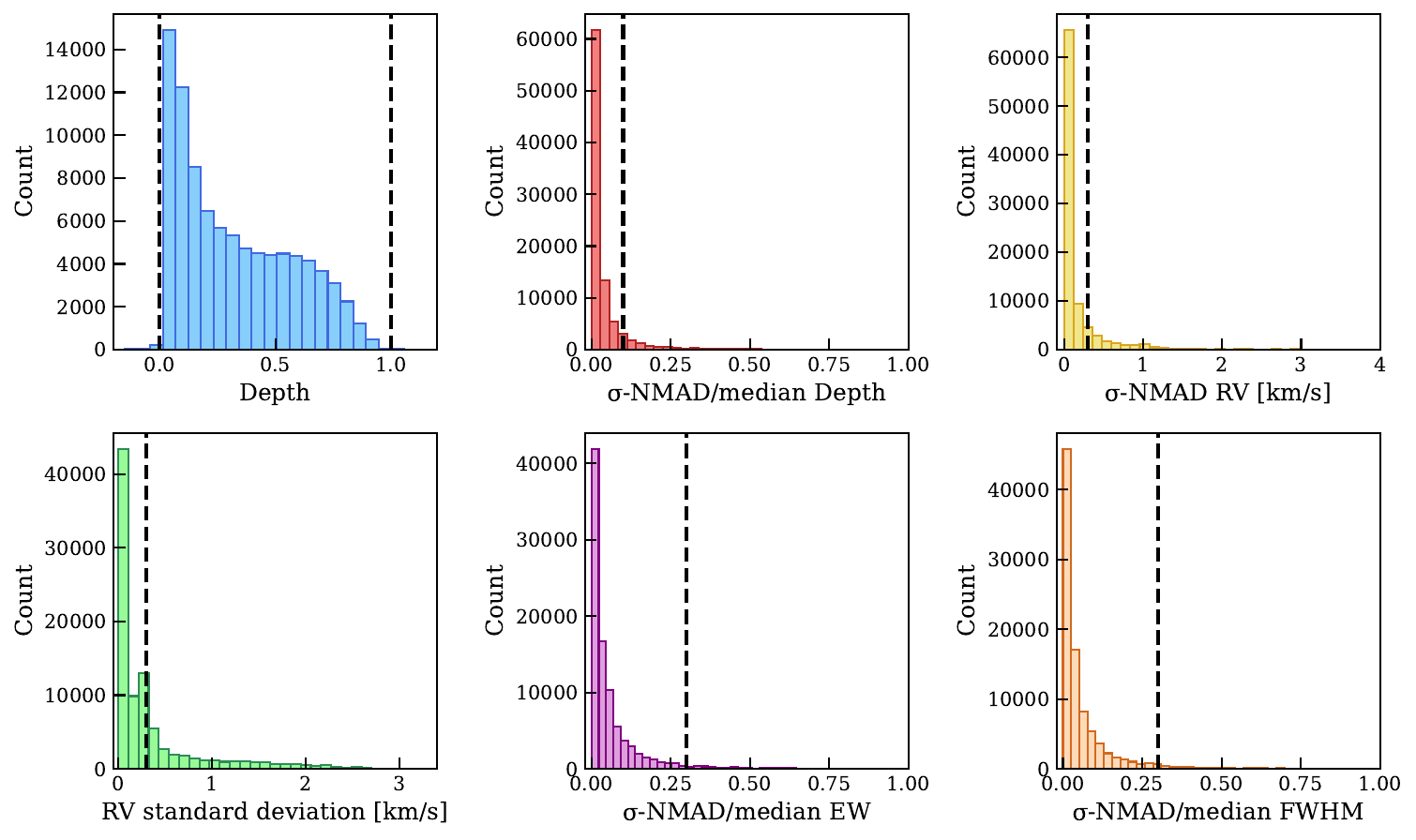}
   \caption{Distribution of the measured and derived properties of the final set of absorption lines used in the analysis. The panels show the depth, the $\sigma_{NMAD}/median$ values (computed as (1.4826 × MAD)/median) for depth, EW, FWHM, and the $\sigma_{NMAD}$ values (computed as 1.4826 × MAD) of the RV, as well as the standard deviation ($\sigma$) of the RV measurements across observations. Vertical dashed lines indicate the reference thresholds adopted during the line-selection process (0 and 1 for depth, 0.1 for the $\sigma_{NMAD}/median$ of the depth, and 0.3 for the other parameters).}

              \label{histograms}%
    \end{figure*}

Since \texttt{TILARA} was originally intended to work with future PoET observations of the Sun, the code includes a list of reference lines derived from 5 solar ESPRESSO UHR spectra reflected from Vesta from \cite{Vardan2020} (hereafter "solar lines"). The solar observations were first corrected for telluric contamination using \texttt{Molecfit} \citep{Smette2015,Smette2020}. 
\texttt{Molecfit} generates a synthetic telluric transmission spectrum by fitting molecular absorption lines using a line-by-line radiative transfer model, taking into account the atmospheric conditions at the time of the observation and the instrumental line spread function. In this work, version~1.5.9 of \texttt{Molecfit} was applied to the ESPRESSO spectra to remove H$_2$O, O$_3$ and O$_2$ features as seen in Fig.~\ref{telluric}, and the resulting transmission model was used to correct the observed spectra prior to the \texttt{ARES} analysis.
The observations were subsequently corrected for the barycentric Earth radial velocity (BERV) and for the measured stellar RVs, ensuring that the spectra were expressed in the Earth’s rest frame. We then ran \texttt{linesearcher} on the corrected spectra to obtain an initial approximation of the line positions.

In the \texttt{linesearcher} step, we tested different parameter combinations for the key variables described in Appendix~\ref{app:line_selection}, while keeping the wavelength range fixed to cover the entire spectrum. We ultimately adopted the following settings for all observations:
\begin{quote}
\texttt{lambdai=3700$\AA$; lambdaf=7900$\AA$; smoothder=8; rejt=0.998\footnote{While the \texttt{rejt} can be computed in different ways (see Appendix~\ref{rejt}), fixing it is sufficient in this case for our purposes, as we are primarily interested in determining the line centers. The line strength (e.g., EW), which is more sensitive to the choice of \texttt{rejt}, is not critical for this analysis.}; lineresol=0.07}
\end{quote}
Where the \texttt{lambdai} and \texttt{lambdaf} parameters determine the studied wavelength range, the \texttt{smoothder} parameter is a smoothing factor, \texttt{rejt} is a continuum normalization parameter, and \texttt{lineresol} determines the minimum allowed separation between two detected lines.
With these parameters, the number of detected lines per observation ranged between 7395 and 7824. This variation is primarily driven by differences in S/N and local continuum placement between observations, which affect the detectability of shallow or blended features close to the detection threshold. Since \texttt{linesearcher} is intentionally applied in a permissive way at this stage, marginal features may be detected in higher-S/N spectra but missed in noisier ones, or treated differently depending on local blending conditions. This step is therefore designed to favor completeness over stability, ensuring that no potentially useful lines are missed. The subsequent \texttt{ARES} refinement and consistency requirements (later in this section) then remove unstable or poorly defined features, resulting in a much more homogeneous and robust final line list.

This preliminary list is then provided to \texttt{ARES}, along with the solar observations themselves, to produce a refined solar line list. For the subsequent \texttt{ARES} step, we again tested various parameter combinations based on the recommendations of \cite{Sousa2007} and \cite{Sousa2015}, and ended up adopting:
\begin{quote}
\texttt{lambdai=3770$\AA$; lambdaf=7900$\AA$
; smoothder=14; space=5$\AA$; rejt=0.995; lineresol=0.07; miniline=1; rvmask='0,0'}
\end{quote}

In this work the value for the parameter \texttt{space} -- which defines the wavelength window around the line center for the fitting process -- is kept fixed for all lines. Although the Doppler width of spectral lines increases toward longer wavelengths, this fixed interval is sufficiently wide to capture the full profile of lines across the entire ESPRESSO range. In the blue part of the spectrum, where line density is high, this window is particularly important as it provides  \texttt{ARES} with enough local context to more accurately resolve blended features and determine the local continuum level in this regard, we note that\texttt{ARES} does not guarantees that all fitted features correspond to single physical lines. The method is fully data-driven and relies solely on the observed spectra. As a result, the final line list may include unresolved blends or composite features, which are treated as effective Doppler tracers and later mitigated statistically with \texttt{TILARA} based on their RV stability.

On another note, the parameter \texttt{rvmask} defines how \texttt{ARES} aligns the reference line list with the observed spectrum. In this context, a 'mask' refers to a user-defined set of spectral features used to estimate the global Doppler shift of the observation. There are two primary options for this alignment: a fixed value for the RV to shift the wavelengths, or an alignment mask containing a few strong, well-defined lines\footnote{Details and recomentations regarding this parameter can be found in \cite{Sousa2015}.}. In this case, the \texttt{rvmask} was set to 0 since the spectra was previously corrected for the BERV and the stellar RV. Additionally, we adopted a slightly lower \texttt{rejt} and a bigger \texttt{smoothder} values in \texttt{ARES} to be more conservative, ensuring that only well-defined and reliably measured lines are retained.

With these settings, \texttt{ARES} measured between 5839 and 6205 lines per observation. The reduction of roughly 1500 lines compared to the \texttt{linesearcher} step reflects the fact that some \texttt{linesearcher} detections were not fitted by \texttt{ARES}, often because the continuum placement defined by the \texttt{rejt} parameter caused weak features to be interpreted as noise rather than genuine lines. While \texttt{linesearcher} was applied more liberally to avoid missing any potential lines, \texttt{ARES} was used to select only robust, reliably detected features.

Once the \texttt{ARES}\footnote{ARES does not make use of external line databases (e.g. VALD) and does not enforce any physical identification of spectral lines. Its role is to identify and fit absorption features present in the data, using Gaussian components whose number and parameters are determined by the local spectral morphology.} results were obtained, we performed an additional refinement step to select a high-confidence set of solar lines. We first select the lines present in at least 3 out of the 5 observations, allowing a tolerance of 0.03~\AA\ in the central wavelength to account for measurement error. This accounted for 5077 lines in total. Then, we computed the line-by-line RVs with the classical Doppler formula using the line positions obtained with \texttt{linesearcher} as the reference wavelengths. We also calculated several statistical descriptors of the line properties across the five solar observations, namely the mean, standard deviation ($\sigma$), variance, median, median absolute deviation (MAD) -- which is less sensitive to outliers --, relative dispersion of the mean, relative dispersion of the median, and the $\sigma_{NMAD}$. These quantities were evaluated for the EW, EW error, FWHM, depth, and RV of each line.

Following robust statistics principles, we adopted the $\sigma_{NMAD}$, defined as
\[
\sigma_{\mathrm{NMAD}} = 1.4826 \times \mathrm{MAD},
\]
as a estimator of the standard deviation, which is less sensitive to outliers than the classical variance\footnote{For a Gaussian distribution without outliers, $\sigma_{\mathrm{NMAD}} = \sigma$.}. For the RV statistics, the $\sigma_{\mathrm{NMAD}}$ value provides a direct measure of the line-to-line stability across the solar observations. For the Depth, EW, and FWHM parameters, we opted to use the relative metric $\sigma_{\mathrm{NMAD}}/\mathrm{median} = (1.4826 \times \mathrm{MAD}) / \mathrm{median}$, which expresses the dispersion normalized by the median value, therefore quantifying their variability in a scale-independent way.

Based on these diagnostics, we applied the following selection thresholds:
\begin{itemize}
    \item $Depth$ between 0 and 1\footnote{A small number of fitted components exhibit non-physical depth values (< 0 or > 1), which are associated with rare multi-Gaussian fits in noisy regions and do not affect the RV measurements.}.
    
    \item  $Depth_{\sigma_{\mathrm{NMAD}\textbf{/median}}} < 0.1$
    
    \item $RV_{\mathrm{\sigma}} < 0.3\ \mathrm{km\,s^{-1}}$
   
    \item  $RV_{\sigma_{\mathrm{NMAD}}} < 0.3\ \mathrm{km\,s^{-1}}$
    
    \item $EW_{\sigma_{\mathrm{NMAD}\textbf{/median}}} < 0.3\ \mathrm{m\AA}$
    
    \item $FWHM_{\sigma_{\mathrm{NMAD}\textbf{/median}}} < 0.3\ \mathrm{\AA}$
\end{itemize}

After applying these constraints, a total of 4454 lines were retained. Some of these parameters are strongly correlated, and therefore certain thresholds may dominate the selection more than others.

The selection criteria were determined empirically from the inspection of the corresponding distributions (Fig.~\ref{histograms}) rather than through formal statistical optimization. Thresholds were set to exclude the extreme tails while retaining the majority of reliable measurements, thereby minimizing the influence of poorly determined lines without imposing overly restrictive cuts. The resulting distributions, shown in Fig.~\ref{histograms} with the adopted thresholds indicated by vertical dashed lines, demonstrate that the retained lines form a tightly clustered and statistically stable sample suitable for line-by-line RV extraction.

\subsection{Step 2: Line characterization with \texttt{ARES}} \label{ARES}

Once a list of reference lines has been identified with \texttt{linesearcher} and refined with \texttt{ARES} (in the demonstration in this paper, using the solar spectra), the next step is to input them together with the spectral observations of the target star into \texttt{ARES} to characterize their absorption lines.

After running \texttt{ARES} on all observations of the star to be studied, we keep only the lines detected in every spectrum. We cross-match the reference–wavelength column across all \texttt{ARES} output files and select the intersection of lines common to all epochs. This ensures that the final line list used for the RV computation is consistent across the full time-series.

\texttt{ARES} internally operates on a local spectral chunk defined by the space parameter, performs a local normalization, and identifies absorption features using derivative-based criteria before fitting Gaussian profiles (for further details refer to \citealt{Sousa2007,Sousa2015}). These observation-dependent windows are not used by \texttt{TILARA}.

In our workflow, we retain the reference wavelengths and the observed line centers as the main inputs for the Doppler shift analysis described in Sect.~\ref{RVanderrors}. Further technical details on the input parameters and output of \texttt{ARES} are provided in Appendix~\ref{app:ARES}.

\subsection{Step 3: RV and RV error calculation} \label{RVanderrors}

The RV measurement of each line \textbf{$l$} is computed using the classical Doppler formula:
\begin{equation}
 \small   RV_l = \frac{\lambda_{\mathrm{obs},l} - \lambda_{\mathrm{ref},l}}{\lambda_{\mathrm{ref},l}} \cdot c
\end{equation}
where $\lambda_{\mathrm{ref}}$ is the reference wavelength\textbf{s} from the  reference line list (common to all observations)  constructed in Section \ref{Solarlines} from the homogenized set of absorption lines consistently detected across the solar spectra; $\lambda_{\mathrm{obs}}$ is the observed wavelength of the line for a given epoch as measured by \texttt{ARES}; and $c$ is the speed of light.

Different approaches to computing the RV error are found in the literature. In this work, we tested four different commonly used methods (see Appendix~\ref{appendixA} for details) and, after comparison, adopted the formalism of \cite{Bouchy2001}. This method uses the wavelength, flux, and flux error from the observations, together with the wavelength of the line center from \texttt{ARES}, to derive the RV error for each line. The RV uncertainty of a single line is given by:
\begin{equation}
\sigma_{RV,l} = \frac{1}{\sqrt{\sum\limits_{i} \frac{1}{\sigma_{\nu, i}^2}}}
\end{equation}
where
\begin{equation} \label{eq3}
\sigma_{\nu, i} = \left| \frac{dF}{d\lambda} \right|^{-1}_i \cdot \frac{c}{\lambda_{i}} \cdot \sigma_{F_i}
\end{equation}
Here, $\sigma_{\nu, i}$ is the velocity uncertainty at each sampled spectral point $i$ within the line, $\frac{dF}{d\lambda}$ is the local derivative of the flux with respect to wavelength, $\lambda_i$ is the wavelength at point $i$, and $\sigma_{F_i}$ is the flux uncertainty at that point provided by the ESPRESSO data reduction pipeline and stored in the corresponding error array of the extracted spectra. Equation~\ref{eq3} is evaluated using the extracted (non-normalized) S1D flux values, since the flux in the S1D files is not continuum-normalized.

To define the wavelength interval comprised by each line, we use the first and second derivatives of the flux with respect to wavelength, similar to \cite{Cretignier2020}. The algorithm searches from the line center outwards:
\begin{itemize}
    \item For the left edge, it steps backwards from the center of the line until it finds the closest point where the first derivative is positive (rising slope) and the second derivative is negative (downward curvature). Marking the start of the line.
    \item For the right edge, it steps forward from the center of the line until it finds the closest point where the first derivative is negative (falling slope) and the second derivative remains negative, marking the end of the line.
\end{itemize}
   \begin{figure*} 
   \centering
   \includegraphics[width=.75\textwidth]{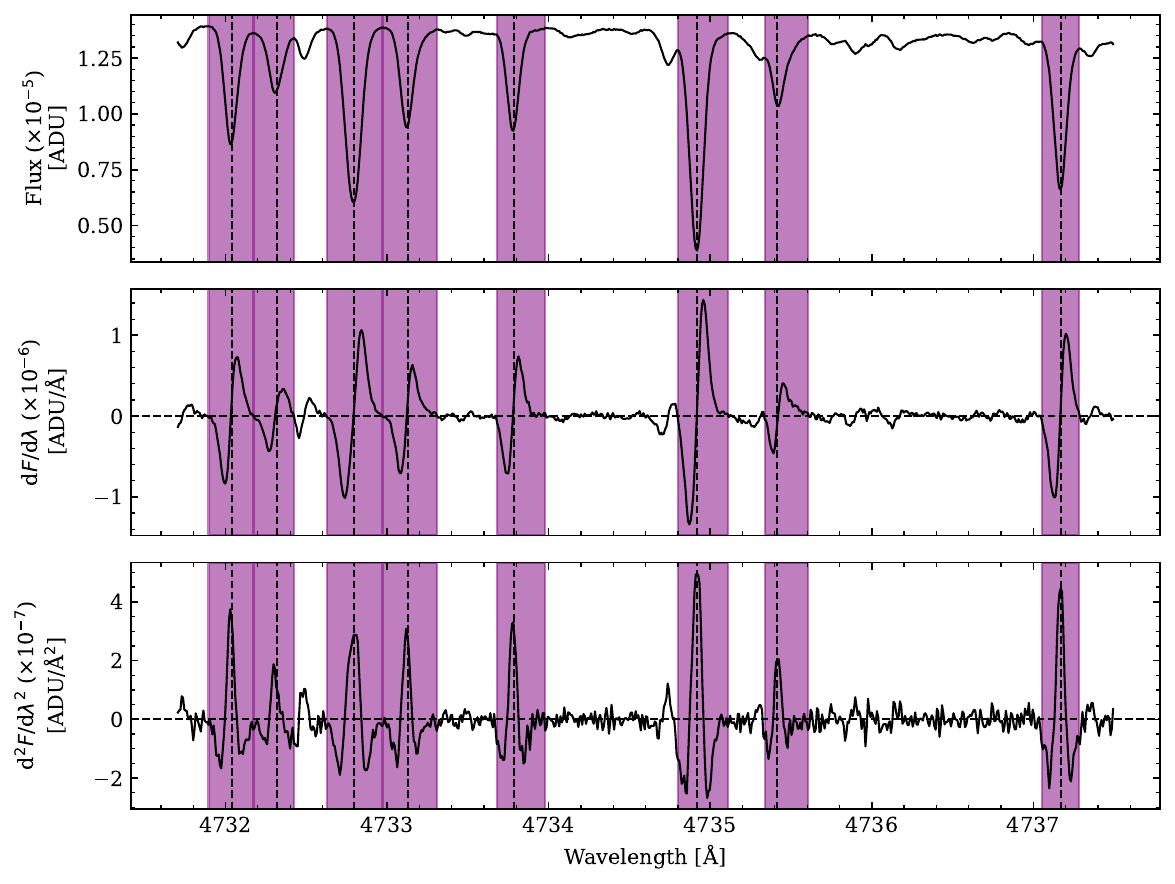}
   \caption{Top: Part of the spectrum of one observation, with the reference line centers (from the solar line list) marked as dashed lines and the adopted line windows shown in purple. The observed line centers are slightly shifted relative to the solar reference due to differences in convective blueshift and gravitational redshift between the Sun and HD~102365. Middle: First derivative of the flux ($dF/d\lambda$), used to locate the extrema that define the window limits. Bottom: Second derivative ($d^{2}F/d\lambda^{2}$), used to confirm the line minima ($d^{2}F/d\lambda^{2} < 0$). The master spectrum includes 2738 lines.}

              \label{Line_windws}%
    \end{figure*}
These gradient-based criteria adapt the measurement window to the actual line profile, ensuring that the calculation captures the full absorption feature without including excessive continuum or neighboring lines as seen in Fig.~\ref{Line_windws}. We emphasize that the derivative-based criteria described here are used only to define the wavelength interval over which the RV uncertainty is computed, after the line center has already been identified via Gaussian fitting with \texttt{ARES}. They are not used to define the spectral lines themselves nor to determine RV.

This procedure cannot be applied independently to each individual spectrum due to variations in S/N across the dataset. For the same spectral line in different observations with different S/N, the code would select a slightly different number of points to compute $\sigma_{\nu,i}$ within the wavelength interval comprised by each line, that also would vary between observations as a consequence of these S/N differences. To overcome this issue, we constructed a master spectrum by taking the median flux across all observations. The wavelength interval comprised by each line and the gradient $\left| \frac{dF}{d\lambda} \right|^{-1}_i$ were then determined from this master spectrum. These values were subsequently used in Equation~\ref{eq3}, ensuring that both the wavelength interval comprised by each line and the gradient ($\left| \frac{dF}{d\lambda} \right|^{-1}_i$) remained fixed across all observations. Finally, to account for S/N differences between the individual observations and the master spectrum (from which the gradient is now obtained, therefore being $\left| \frac{dF}{d\lambda} \right|^{-1}_{i,master}$ from now on), we scale the gradient with the median ratio between the master spectrum's flux and the current observation's. The modified equation \ref{eq3}, becomes the following expression:
\begin{equation}
\sigma_{\nu, i} = \left| \frac{dF}{d\lambda} \right|^{-1}_{i,master} \cdot \mathrm{median}\left( \frac{F_{master}}{F_{obs}}\right)\cdot \frac{c}{\lambda_{i}} \cdot \sigma_{F_i}
\end{equation}
Here, $\mathrm{median}\left( \frac{F_{master}}{F_{obs}}\right)$ denotes a single global scaling factor computed over the full spectral range, and not a wavelength-dependent quantity. This global scaling does not explicitly correct for chromatic flux variations across the spectrum (e.g. due to airmass-dependent effects), but $\sigma_{Fi}$ remains proportional to the local S/N of the observed spectrum. Future versions of the pipeline could incorporate wavelength-dependent flux scaling to better account for chromatic effects.
This master spectrum ensures that the windows of the spectral lines, and the gradients inside these are well calculated regardless of the level in S/N that the different observations might have.

\subsection{Step 4: Weighted RV time series}\label{weighting}

When working with line-by-line RV measurements, some lines can be affected by photon noise, spectral blending, telluric contamination, and/or instrumental artifacts. Such lines may produce RV outliers that can bias the final results if they are treated equally to well-behaved lines. 

It should be noted that the RV precision of an individual spectral line is generally not sufficient to identify low-amplitude stellar activity (at the m/s level) as a statistical outlier within a single epoch. However, lines that are systematically sensitive to such effects will manifest a higher global standard deviation over the course of an observing campaign. Such lines produce RV outliers or noisy contributions that can bias the final results if they are treated equally to well-behaved lines.

To address this, \texttt{TILARA} implements two different independent approaches: iterative sigma-clipping and down-weighting.

\subsubsection{Sigma-clipping}\label{sigma_clip}

Sigma-clipping is applied in \texttt{TILARA} to the distribution of line-by-line RVs within each individual observation, prior to computing the final RV time-series. The procedure consists of two steps. First, \texttt{TILARA} applies sigma-clipping to the RV error to remove lines whose error estimates are inconsistent with the RV error overall distribution. Then, from the remaining lines, the code performs a second sigma-clipping on the standard deviation of the RVs for each line across all observations, removing lines that show anomalously large intrinsic variability. This two-stage filtering ensures that only reliable spectral lines contribute to the final weighted RV for each observation.

In \texttt{TILARA}, both the sigma threshold and the number of iterations are configurable by the user, allowing the procedure to be tuned according to the noise characteristics of the dataset.  

Once the sigma-clipping step is complete, the final RV for an observation is computed as a weighted mean of the remaining line-by-line RVs:
\begin{equation}
\small RV_{\mathrm{weighted}} = \frac{\sum\limits_{l} \frac{RV_l}{\sigma_{RV,l}^2}}{\sum\limits_{l} \frac{1}{\sigma_{RV,l}^2}}
\end{equation}
where $RV_l$ and $\sigma_{RV,l}$ are the RV and RV error for line $l$, respectively. The corresponding weighted uncertainty is given by:
\begin{equation}
\sigma_{RV,\mathrm{weighted}} =
\frac{1}{\sqrt{{\sum\limits_{l} \frac{1}{\sigma_{RV,l}^2}}}}
\end{equation}

This inverse-variance weighting ensures that measurements with smaller error contribute more to the final RV, while noisier measurements are naturally down-weighted.

\subsubsection{Down-weighting with Lorentzian fit}\label{Lorentz}

The down-weighting strategy implemented in \texttt{TILARA} is conceptually inspired by the probabilistic approach of \citet{Artigau2022}, which models the distribution of residual RVs obtained from template-matching to identify and down-weight potential outlier spectral features. However, an important distinction arises from the fact that in our case, the RVs of individual absorption lines are obtained from \texttt{ARES} by fitting Gaussian profiles. This has a direct consequence for the statistical behavior of the line-by-line RVs. Therefore, unlike template-matching approaches, \texttt{TILARA} does not assume that each line's RV time series is centered around zero. Each line is allowed to have its own intrinsic constant RV offset, reflecting physical line-dependent effects. When RVs are computed through a typical template matching algorithm, the template itself defines the rest-frame wavelength scale. As a result, the distribution of RVs per line is naturally centered around zero by construction, and deviations primarily reflect temporal variability. The down-weighting scheme of \citet{Artigau2022} therefore operates on line-by-line RV residuals, assuming a symmetric distribution around zero.

\begin{figure*}
   \centering
   \includegraphics[width=0.75\textwidth]{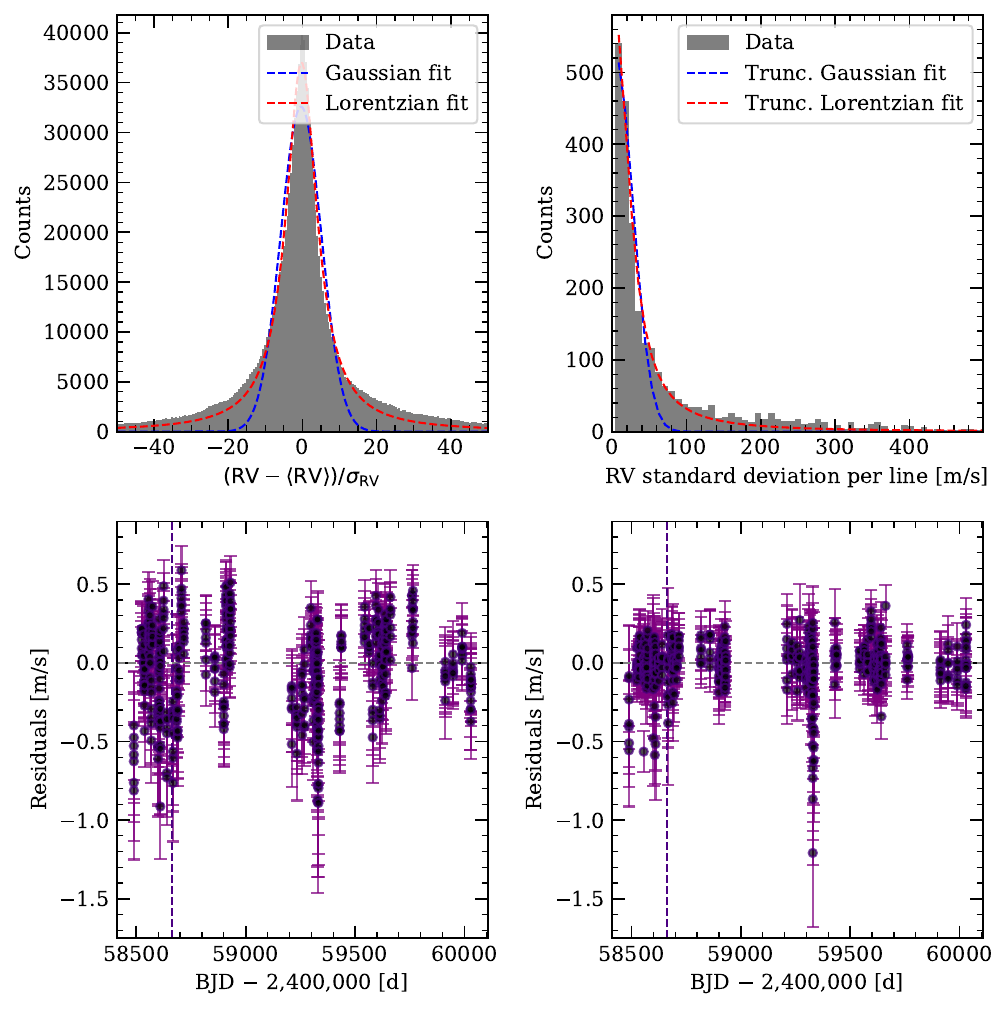}
   \caption{Summary of the down-weighting procedures tested for \texttt{TILARA} (for a detailed description of them see Appendix~\ref{down-weight-tests}). \textit{Top left panel}: Distribution of the RV values (mean-subtracted and $\sigma$-normalized) per observation, together with the best-fit Gaussian (blue) and Lorentzian (red) models. \textit{Top right panel}: Distribution of the RV standard deviation per line, showing the histogram (gray) and the corresponding truncated Gaussian and truncated Lorentzian fits. \textit{Bottom left panel}: Residual RV time-series obtained by subtracting the RVs derived using a Lorentzian fit from those derived using a Gaussian fit (mean-subtracted and $\sigma$-normalized) per observation. The residuals are centered around zero with comparable scatter. \textit{Bottom right panel}: Residual RV time-series obtained by subtracting the RVs derived using a truncated Lorentzian fit from those derived using a truncated Gaussian fit when modeling the distribution of the RV standard deviation per line. The residuals are centered around zero with comparable scatter. A vertical black line marks 27 June 2019, indicating the transition from ESPRESSO18 to ESPRESSO19 on both bottom panels.}

              \label{RV/RVerror}%
    \end{figure*}

In contrast, when fitting Gaussians to individual absorption lines, each line is compared to a reference wavelength (in this paper the wavelengths derived using \texttt{ARES} on the solar spectrum). Small errors in these reference wavelengths, unresolved blends, profile asymmetries, and instrumental or stellar velocity offsets -- due to differences in the physical properties of the target star and the reference star used to define the initial line-list -- introduce line-dependent constant shifts that persist across all epochs. Therefore, the line-by-line RVs calculated with lines extracted with \texttt{ARES} are not centered at zero and cannot be directly modeled using the same formalism as in \citet{Artigau2022} as seen in the \textit{Top left panel} of Fig.~\ref{RV/RVerror}. We tested several alternatives to re-center the RV distributions at the measurement level, but these approaches were found to be noise-sensitive and did not provide a stable basis to identify unreliable spectral lines (for a more detailed description of the different tests, see Appendix \ref{down-weight-tests}).

Instead, we find that a more robust diagnostic of line quality is the temporal stability of each line's RV. For each line $\ell$, we compute the standard deviation of its median-centered RVs across all observations:
\begin{equation}
\sigma_{\ell} = \mathrm{std}\bigl(\mathrm{RV}_{\ell}(t))
\end{equation}

We tested subtracting the median RV of each line as an alternative normalization, but found through injection–recovery tests that this procedure partially absorbs coherent RV signals, including planetary signatures.

 The empirical distribution of the standard deviations exhibits a pronounced peak and a heavy tail corresponding to lines strongly affected by stellar activity, blends with other stellar lines or telluric residuals, or possible instrumental systematics (see the \textit{Top right panel} of Fig.~\ref{RV/RVerror}).

To model this distribution, we opted to fit a truncated Lorentzian function:
\begin{equation}
L(x) = \frac{A}{1 + \left(\frac{x - x_{0}}{\gamma}\right)^{2}} \,,\, x \in [x_0, \infty)
\end{equation}
where $x$ is the normalized dispersion, $A$ the amplitude, and $\gamma$ the half-width at half-maximum. The location $x_{0}$ is fixed to the minimum observed dispersion, corresponding to the most stable lines. 

Each spectral line is then assigned a weight
\begin{equation}
p_{\ell} = L(\sigma_{\ell}),
\end{equation}
to penalize lines that exhibit large RV standard deviations across observations. This can be seen in the Truncated Lorentzian fit, represented in red in the \textit{Top right panel} of Fig.~\ref{RV/RVerror}, where the histogram (gray) shows the empirical spread of temporal RV variability across lines, after median-centering each line. The red curve represents the best-fit truncated Lorentzian model, where the center is fixed to the minimum observed dispersion. Lines near the peak correspond to the most stable Doppler tracers, while those in the heavy tail display larger variability and are consequently down-weighted in the final RV computation. Then, the weights $p_{\ell}$ are normalized such that $\sum_{\ell} p_{\ell} = N_{\mathrm{lines}}$. The final epoch-wise RV is computed as a weighted inverse-variance mean:
\begin{equation}
RV_{\mathrm{weighted}} = \frac{\sum\limits_{l} \frac{p_l \, RV_l}{\sigma_{RV,l}^2}}{\sum\limits_{l} \frac{p_l}{\sigma_{RV,l}^2}}
\end{equation}
with associated uncertainty
\begin{equation}
\sigma_{RV,\mathrm{weighted}} = \frac{1}{\sqrt{\sum\limits_{l} \frac{p_l}{\sigma_{RV,l}^2}}}
\end{equation}
This approach retains the contribution of all lines while naturally down-weighting those whose RVs exhibit inconsistent temporal behavior, improving robustness relative to hard rejection schemes and reducing the impact of blended or activity-sensitive lines on the final RV time series.

The Lorentzian fit to the distribution of the RV standard deviation per line is used as an empirical and robust way to characterize the observed population of stable and unstable lines. We do not assume that this distribution is intrinsically Lorentzian, nor that the same shape should apply to other datasets. This approach provides a smooth down-weighting of lines with large temporal RV scatter and avoids an overly aggressive weighting.

\section{HD~102365: Application}\label{HD~102365}
As previously mentioned, the code was originally developed for the study of disk-resolved solar observations to be carried out with the soon to come PoET. Thus, although it can in principle be applied to any stellar spectral type, for demonstration purposes in this paper we use the list of solar spectral lines described in Sect. \ref{Solarlines}, and we test the code on a Sun-like star to illustrate its performance.

We tested the performance of \texttt{TILARA} on 520 HR ESPRESSO observations\footnote{No explicit telluric correction (beyond the O$_2$ correction applied by the ESPRESSO DRS) was applied to the target spectra in this work.} of the bright (V=4.89; \citealt{Figueira2025}) Sun-like star HD~102365. The star's spectral type that has been discussed to be G5 \citep{Tinney2011}, G8 \citep{Hojjatpanah2019},  and more recently G6V by \cite{Figueira2025}. HD 102365 is a low activity star, with reported values of $log R'_{HK}=-4.99$ \citep{Tinney2011} and $log R'_{HK}=-4.88 \pm 0.01$ \citep{Hojjatpanah2019}, and it presents a rotational period of 40 days \cite{Noyes1984}.

HD 102365 was selected as a test case primarily because it is a bright, RV-quiet, slowly rotating G-type star with a well-characterized spectrum and a long time series of publicly available high-resolution ESPRESSO observations. These properties make it a convenient and realistic benchmark for validating the \texttt{TILARA} methodology, in particular for performing injection–recovery tests and comparing line-by-line RVs with standard CCF-based measurements. We note that \texttt{TILARA} was originally developed in the context of forthcoming solar observations with PoET, and Sun-as-a-star data would indeed represent an ideal validation case. However, high-quality Sun-as-a-star spectra obtained with instrumental setups directly comparable to ESPRESSO are currently limited, and PoET observations are not yet available. HD 102365 therefore serves as a close proxy for demonstrating the performance of the method on a Sun-like target under realistic observational conditions. The reference line list used in this work was constructed from a high signal-to-noise solar spectrum rather than from HD 102365 itself. This choice was made to ensure a stable, high-quality reference that is minimally affected by stellar variability and to avoid imprinting star-specific RV signals or line-profile variations into the reference. While the spectral classification of HD 102365 varies slightly in the literature (ranging from different G subtypes), this difference does not affect the methodological demonstration presented here. Building fully star-specific reference line lists is a natural next step for future applications of \texttt{TILARA} and might be explored in subsequent work.

In this work, we specifically use the S1D FITS files, which are one-dimensional spectra produced by rebinning and merging the extracted S2D echelle orders onto a uniform wavelength grid in the solar system barycentric frame. While this resampling process conserves flux, it introduces correlations between adjacent pixels that should be accounted for in detailed quantitative analyses (see the ESPRESSO DRS pipeline manual\footnote{\href{https://ftp.eso.org/pub/dfs/pipelines/espresso/espdr-pipeline-manual-1.0.0.pdf}{ESPRESSO DRS Pipeline Manual v1.0.0}}).

The reference line list used (described in Section~\ref{Solarlines}) was constructed from ESPRESSO UHR observations. This choice was motivated by the original context in which \texttt{TILARA} was developed, namely future disk-resolved solar observations with PoET, which will operate with ESPRESSO in UHR mode, and by the availability of high-quality solar spectra in this configuration. The HD~102365 observations analyzed here were obtained in HR mode.The difference in spectral resolution between HR and UHR does not represent a limitation for the applicability of \texttt{TILARA}. For Sun-like stars, the widths of absorption lines are dominated by intrinsic stellar broadening mechanisms such as rotation, macroturbulence, and convective motions, such that the HR–UHR resolution difference has a negligible impact on the identification of individual spectral lines \citep{Vardan2020}. Since \texttt{TILARA} relies on reference line centers rather than detailed line profiles, UHR observations are not a requirement for RV extraction. The use of UHR data therefore represents a conservative choice for constructing a high-quality reference line list, rather than a necessity imposed by the method.

Even though this star was proposed to host a Neptune-mass planet \citep{Tinney2011}, the work from \cite{Figueira2025}, as well as ours (see Section \ref{periodogram}) show that the planet is not found with ESPRESSO data (in our analysis, at least without the modeling of stellar activity). The observations have a median S/N of 267 and span from 5 January 2019 to 26 March 2023, providing a long-term dataset ideal for testing the stability and performance of our code. Regarding the temporal span of the data set, on the 27th of June 2019 the ESPRESSO spectrograph underwent a physical intervention, that lead to different instrumental profiles before (ESPRESSO 18) $\&$ after (ESPRESSO 19), producing a jump in the data that will later be seen in Figs. \ref{TILARA} and \ref{residuals}. This intervention required that we treated the data as two different datasets when doing the final comparisons. 

Our evaluation focused on three key aspects: (i) the impact of different values for the \texttt{rejt} parameter in \texttt{ARES}, which is fully explained in Appendix~\ref{rejt} , (ii) the effect of various methods for computing RV error (detailed in Appendix \ref{appendixA}), and (iii) the performance of different outlier rejection strategies for the individual lines (detailed in Appendix \ref{down-weight-tests}). Once these tests were performed, and the best approaches for both \texttt{ARES} and \texttt{TILARA} were implemented, we computed the RV time-series for HD~102365. Then we compared our results with the ones from other RV extraction codes, and we tested the robustness of \texttt{TILARA} by doing injection/recovery simulations of planetary signals

\subsection{RV time-series with \texttt{TILARA} and other codes}

Once the reference solar line list was established (see Sect. \ref{Solarlines}), we used \texttt{ARES} to measure the central wavelengths of the absorption lines in the HD~102365 spectra. These values were then used to compute the line-by-line RVs and their associated error, following the procedure described in Sect.~\ref{RVanderrors}. It is to be remarked that, in this step, the parameter \texttt{rvmask} of \texttt{ARES} was set to the systemic velocity of the star $\sim$17.3 [km\,s$^{-1}$], to ensure proper alignment of the reference line list with the observed spectra.

Since \texttt{TILARA} can compute the final RV time-series using two different approaches (described in the sigma-clipping and downweighting sections, Sects.~\ref{sigma_clip} and \ref{Lorentz} respectively), we tested both methods independently for the example presented in this paper, star HD~102365. 

To place our results in context, we also compared them with other extraction algorithms, namely two other line-by-line implementations -- the \texttt{LBL}\footnote{\url{https://lbl.exoplanets.ca}} code from \cite{Artigau2022} and the \texttt{ARVE}\footnote{\url{https://github.com/almoulla/arve}} code from \cite{Almoulla2025} --, the template-matching code s-BART\footnote{\url{https://github.com/iastro-pt/sBART}} from \cite{Silva2022}, and the \texttt{CCF} RVs computed by the official ESPRESSO pipeline.

The \texttt{LBL}, \texttt{ARVE}, and \texttt{s-BART} pipelines all rely on template matching but differ in their implementation. \texttt{LBL} and \texttt{ARVE} are line-by-line methods, whereas \texttt{s-BART} applies a global template-matching approach. Each code adopts its own strategy for outlier rejection or down-weighting of individual measurements.
    
\subsubsection{Impact of sigma-clipping and down-weighting}

To assess the effect of the two outlier-rejection strategies implemented in \texttt{TILARA}, we computed the RV time-series using both the sigma-clipping and down-weighting approaches and compared the resulting measurements. In the down-weighting case, the contribution of each line is scaled according to a truncated Lorentzian distribution fitted to the normalized standard deviation of the line-by-line RVs, as described in Sect.~\ref{Lorentz}. The fitted distribution is shown in the Truncated Lorentzian fit, represented in red in the \textit{Top right panel} of Fig.~\ref{RV/RVerror}, where the peak corresponds to the most stable lines and the extended tail identifies lines that are progressively down-weighted in the final RV estimate.

\begin{figure*}
   \centering
   \includegraphics[width=0.8\textwidth]{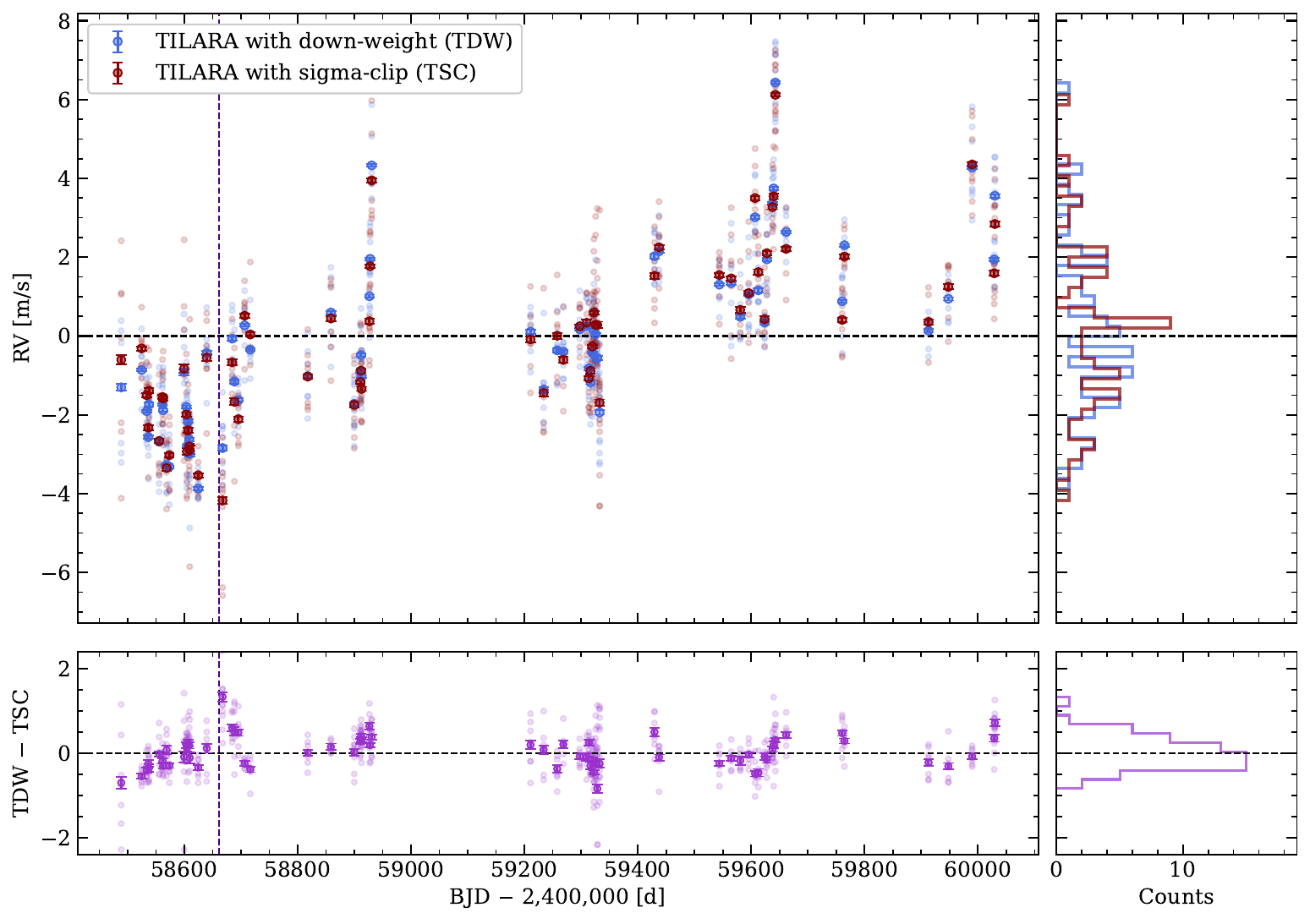}
   \caption{ \textit{Top panel:} Comparison of RV time-series obtained with the \texttt{TILARA} pipeline using the two different outlier-rejection strategies: sigma-clipping (TSC, red) and down-weighting (TDW, blue). The bright-colored points correspond to nightly binned data, while the fainter points in the same colors show the original unbinned measurements. The histogram on the right shows the RV distribution for each method.
\textit{Bottom panel:} Residuals between both methods (TDW – TSC) as a function of BJD, with the associated residual distribution shown in the histogram. This highlights the level of agreement and any systematic differences between the two approaches. A vertical indigo line marks 27 June 2019, indicating the transition from ESPRESSO18 to ESPRESSO19.}

              \label{TILARA}%
    \end{figure*}

A direct comparison between the two \texttt{TILARA} configurations is shown in Fig.~\ref{TILARA}. Both methods recover the same overall RV behavior across the ESPRESSO 18 and ESPRESSO 19 datasets, with the sigma-clipping achieving a standard deviation of 1.01~[m\,s$^{-1}$] for ESPRESSO 18 and 1.90~[m\,s$^{-1}$] for ESPRESSO 19, while the down-weighted series reach a standard deviation of 0.94~[m\,s$^{-1}$] ESPRESSO 18 and 1.87~[m\,s$^{-1}$] for ESPRESSO 19. The down-weighting mode displays slightly reduced scatter, as confirmed by the narrower histogram in Fig.~\ref{TILARA} and the overall values depicted in Table \ref{tab:rv_stats}. The residual panel highlights that the differences between the two approaches are small and mostly random, indicating that both methods are internally consistent and do not introduce significant systematic offsets.

\begin{figure*}
   \centering
   \includegraphics[width=0.78\textwidth]{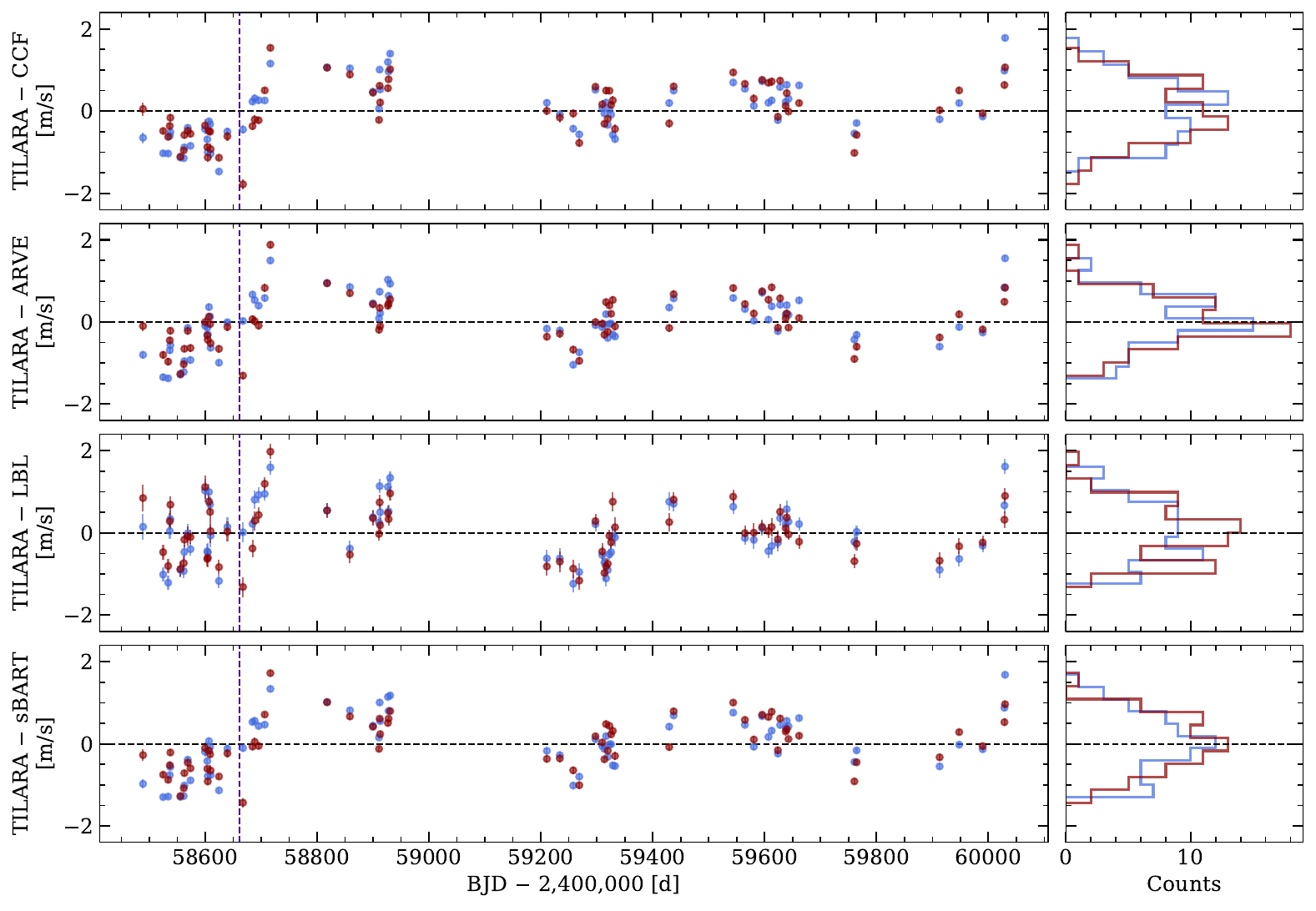}
   \caption{Residual RVs between the two \texttt{TILARA} configurations (down-weighting, TDW, in blue; and sigma-clipping, TSC, in red) when compared to four different RV extraction methods: CCF, ARVE, LBL, and \texttt{sBART} (one panel per method, from top to bottom). The histograms on the right display the distribution of residuals for both \texttt{TILARA} approaches in each comparison, allowing a visual assessment of their relative scatter and systematic differences. A vertical indigo line marks 27 June 2019, indicating the transition from ESPRESSO18 to ESPRESSO19. For consistency with Fig.~\ref{TILARA}, the x–axis limits of the residual panels and both axes of the histograms are kept fixed, enabling a direct visual comparison between the two figures.}

              \label{residuals}%
    \end{figure*}

The performance of \texttt{TILARA} is further evaluated against the four, previously explained independent pipelines in Fig.~\ref{residuals}. The residuals are computed with respect to the two \texttt{TILARA} solutions, allowing a direct visual assessment of their relative precision. All pipelines remain consistent at the $\sim$\,m\,s$^{-1}$ level, but the scatter of the residuals is, in general, smaller when referenced to the down-weighted \texttt{TILARA} RVs. The \texttt{LBL} results show the broadest distributions, consistent with their larger formal error. The results in Table~\ref{tab:rv_stats} confirm that both \texttt{TILARA} configurations reach a level of RV precision comparable to the other state-of-the-art pipelines, for both ESPRESSO 18 and ESPRESSO 19 and in both the unbinned and nightly binned cases. 

Regarding the residuals between methods (see Table \ref{tab:residual_std}), the smallest standard deviation is obtained for the residuals between both \texttt{TILARA} methods and the CCF in ESPRESSO 18, with $\sigma = 0.33~\mathrm{m,s^{-1}}$. While the two \texttt{TILARA} variants show a similar overall scatter to the reference methods, the down-weighted solution systematically yields the lowest standard deviation and the smallest formal error, demonstrating that it maintains competitive precision while providing tighter error bars than the alternative pipelines.

\begin{table*}[t!]
\centering
\caption{Standard deviation and average error bars for HD~102365 obtained from the CCF, \texttt{ARVE}, \texttt{LBL}, \texttt{sBART}, and \texttt{TILARA} methods (including the two outlier removal/down-weighting approaches). Results are shown separately for ESPRESSO 18 and ESPRESSO 19, for both unbinned and nightly binned data.}
\label{tab:rv_stats}
\begin{tabular}{lcccccccc}
\toprule
\multirow{3}{*}{\textbf{Method}} &
\multicolumn{4}{c}{\textbf{Unbinned}} &
\multicolumn{4}{c}{\textbf{Binned by night}} \\
\cmidrule(lr){2-5} \cmidrule(lr){6-9}
& \multicolumn{2}{c}{\textbf{ESPRESSO 18}} & \multicolumn{2}{c}{\textbf{ESPRESSO 19}} &
  \multicolumn{2}{c}{\textbf{ESPRESSO 18}} & \multicolumn{2}{c}{\textbf{ESPRESSO 19}} \\
\cmidrule(lr){2-3} \cmidrule(lr){4-5} \cmidrule(lr){6-7} \cmidrule(lr){8-9}
& \textbf{Std. dev.} & \textbf{Avg. err.} &
  \textbf{Std. dev.} & \textbf{Avg. err.} &
  \textbf{Std. dev.} & \textbf{Avg. err.} &
  \textbf{Std. dev.} & \textbf{Avg. err.} \\
 & [m\,s$^{-1}$] & [m\,s$^{-1}$] & [m\,s$^{-1}$] & [m\,s$^{-1}$] 
 & [m\,s$^{-1}$] & [m\,s$^{-1}$] & [m\,s$^{-1}$] & [m\,s$^{-1}$] \\
\midrule
\texttt{TILARA} (with sigma-clip) & 1.44 & 0.18 & 2.20 & 0.18 & 1.01 & 0.07 &1.90 & 0.06 \\

\texttt{TILARA} (with down-weight) & 1.20 & 0.13 & 2.10 & 0.12 & 0.94 & 0.05 & 1.87 & 0.04 \\
\hline
\texttt{CCF} (DRS) & 1.09 & 0.16 & 1.95 & 0.15 & 0.89 & 0.06 & 1.75 & 0.05 \\
\texttt{ARVE} \citep{Almoulla2025} & 1.23 & 0.16 & 2.04 & 0.16 & 1.05 & 0.06 & 1.83 & 0.06 \\
\texttt{LBL} \citep{Artigau2022} & 1.26 & 0.53 & 2.06 & 0.52 & 1.03 & 0.20 & 1.83 & 0.18 \\
s-BART \citep{Silva2022}& 1.16 & 0.13 &1.96 & 0.13 & 0.98 & 0.05 & 1.77 & 0.05 \\
\bottomrule
\end{tabular}
\end{table*}

\begin{table*}[t!]
\centering
\caption{Standard deviation of the RV residuals  between the two \texttt{TILARA} configurations -- down-weighting (TDW) and sigma-clipping (TSC) -- and the four comparison pipelines, computed separately for ESPRESSO~18 and ESPRESSO~19.}
\label{tab:residual_std}
\begin{tabular}{lcc}
\toprule
 & \multicolumn{2}{c}{\textbf{Standard deviation} [m\,s$^{-1}$]} \\
\cmidrule(lr){2-3}
\textbf{Method Comparison} & \textbf{ESPRESSO 18} & \textbf{ESPRESSO 19} \\
\midrule
\texttt{TILARA} (with sigma-clip) – CCF   & 0.33 & 0.59\\ 
\texttt{TILARA} (with down-weight) – CCF   & 0.33 & 0.55 \\

\midrule
\texttt{TILARA} (with sigma-clip) – ARVE  & 0.38 & 0.56\\
\texttt{TILARA} (with down-weight) – ARVE  & 0.53 & 0.53 
 \\
\midrule
\texttt{TILARA} (with sigma-clip) – LBL   & 0.62 & 0.65\\ 
\texttt{TILARA} (with down-weight) – LBL   & 0.68 & 0.69 \\

\midrule
\texttt{TILARA} (with sigma-clip) – sBART & 0.33 & 0.57 \\
\texttt{TILARA} (with down-weight) – sBART & 0.44 & 0.56 
\\
\bottomrule
\end{tabular}
\end{table*}

\subsection{Testing the robustness of \texttt{TILARA} through injection-recovery simulations}\label{injection}

The injection and recovery of synthetic planetary signals is a standard approach to assess both the detectability of planets in a given RV dataset and the reliability of the RV-retrieval pipeline. Even in cases where no planets are detected, planetary signals may still be present but remain undetected due to stellar activity, instrumental noise, or irregular time sampling. By adding artificial Keplerian signals with known parameters directly into the observed spectra and attempting to recover them with the same analysis framework used for real detections, we can quantify the sensitivity of the dataset and evaluate the performance of our methodology.

In our implementation, the injection was performed by shifting the wavelength of each original spectrum according to the Doppler signal that a planet with certain orbital parameters would induce. This approach ensures that the synthetic signal is introduced in a physically consistent way, directly mimicking the effect that an actual planetary companion would have on the stellar spectrum.

As a first test, we injected a synthetic planet with a period of $P = 100.00$\,days, a semi-amplitude of $K = 10.00\,\mathrm{m\,s^{-1}}$, and an eccentricity of $e = 0.00$ directly into the spectra. The modified spectra were then processed with \texttt{TILARA} following the same steps applied to the original data. The resulting RVs were analyzed with \texttt{KIMA}\footnote{\url{https://github.com/kima-org/kima}} \citep{Faria2018}, treating the ESPRESSO~18 and ESPRESSO~19 observations as independent instruments because of the previously mentioned different instrumental profiles that they exhibit.\texttt{KIMA} is a package for the detection and characterization of exoplanets from RV data, fitting sums of Keplerian signals using Diffusive Nested Sampling, allowing the number of planets to be a free parameter, and modeling stellar activity with Gaussian processes. In this work, we gathered the RVs and their associated errors from each observation, ran \texttt{KIMA} with a single-planet model, and used the resulting posterior samples to infer the planet’s amplitude, period, and phase. We also generated diagnostic plots including the data, phase-folded RV curves, posterior distributions, and parameter corner plots.

\begin{figure*}[htbp]
    \centering
    \begin{subfigure}[b]{0.43\textwidth}
        \centering
        \includegraphics[width=\textwidth]{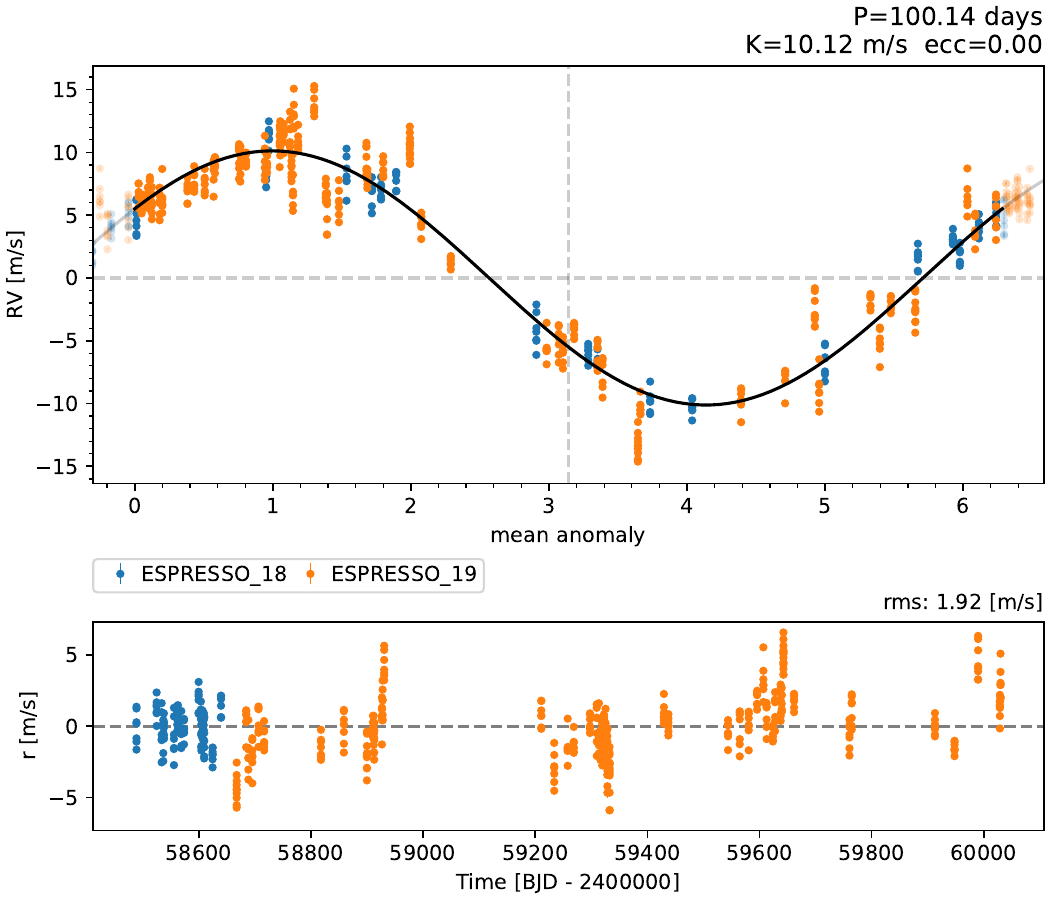}
        \label{KIMA_10}
    \end{subfigure}
    \hfill
    \begin{subfigure}[b]{0.43\textwidth}
        \centering
        \includegraphics[width=\textwidth]{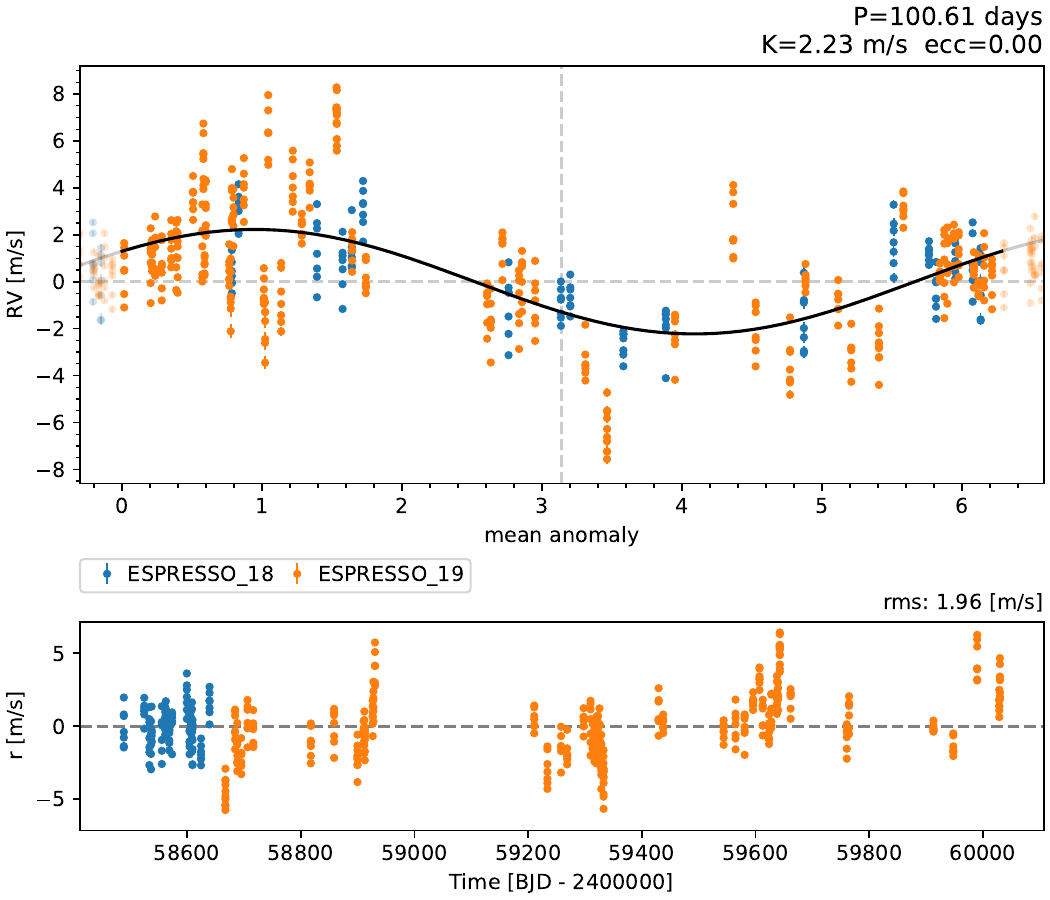}
        \label{KIMA_2}
    \end{subfigure}
    \caption{Phase-folded RV curves recovered for the injected synthetic planets with a period of $100.00$\,days and eccentricity $e = 0.00$, after applying the down-weighting method. 
    The \textit{left panel} shows the recovered RV curve for the injected signal with $K = 10.00\,\mathrm{m\,s^{-1}}$, and the \textit{right panel} for $K = 2.00\,\mathrm{m\,s^{-1}}$. 
    In both cases, the solid line represents the best-fit Keplerian model obtained with \texttt{KIMA}, successfully retrieving the injected signal using the RVs derived with \texttt{TILARA}.}
    \label{KIMA}
\end{figure*}

To perform a more restrictive test, we repeated the experiment using the same period and eccentricity, but reducing the semi-amplitude to $K = 2.00\,\mathrm{m\,s^{-1}}$. In both cases, the injected signal was successfully recovered using both the sigma-clipped and the down-weighted RVs, as summarized in Table~\ref{tab:planet}. Figure~\ref{KIMA} shows the phase-folded RV curves recovered with \texttt{KIMA} for the two injected amplitudes after being treated with the down-weight method.

\begin{table*}[h!]
\centering
\caption{Injected and recovered parameters for two synthetic planetary signals in the HD~102365 RV data.}
\label{tab:planet}
\renewcommand{\arraystretch}{1.3} 
\setlength{\tabcolsep}{8pt} 
\begin{tabular}{lcccc}
\toprule
 & \multicolumn{2}{c}{\textbf{Planet 1: $K=10.00$\,[m\,s$^{-1}$] , $P=100.00$\,[d]}} 
 & \multicolumn{2}{c}{\textbf{Planet 2: $K=2.00$\,[m\,s$^{-1}$] , $P=100.00$\,[d]}} \\
\cmidrule(lr){2-3} \cmidrule(lr){4-5}
 & \textbf{Sigma-clipping} & \textbf{Down-weight} 
 & \textbf{Sigma-clipping} & \textbf{Down-weight} \\
\midrule
\textbf{Recovered $K$ [m\,s$^{-1}$]} 
 & $10.12^{\,+0.11}_{\,-0.14}$ & $10.12^{\,+0.11}_{\,-0.10}$ 
 & $2.46^{\,+0.28}_{\,-0.30}$ & $2.23^{\,+0.11}_{\,-0.14}$ \\
\textbf{Recovered $P$ [d]} 
 & $100.12^{\,+0.06}_{\,-0.03}$ & $100.14^{\,+0.06}_{\,-0.03}$ 
 & $99.86^{\,+0.39}_{\,-0.35}$ & $100.61^{\,+0.19}_{\,-0.14}$ \\
\bottomrule
\end{tabular}
\end{table*}

\subsection{RV periodogram of HD~102365} \label{periodogram}

Previous studies reported the presence of the Neptune-mass planet HD~102365 b \citep{Tinney2011}, so to accurately interpret the results obtained with \texttt{TILARA}, we computed a periodogram from our RV results of HD~102365 to asses the existence of this planet. 
   \begin{figure*}
   \centering
   \includegraphics[width=0.7\textwidth]{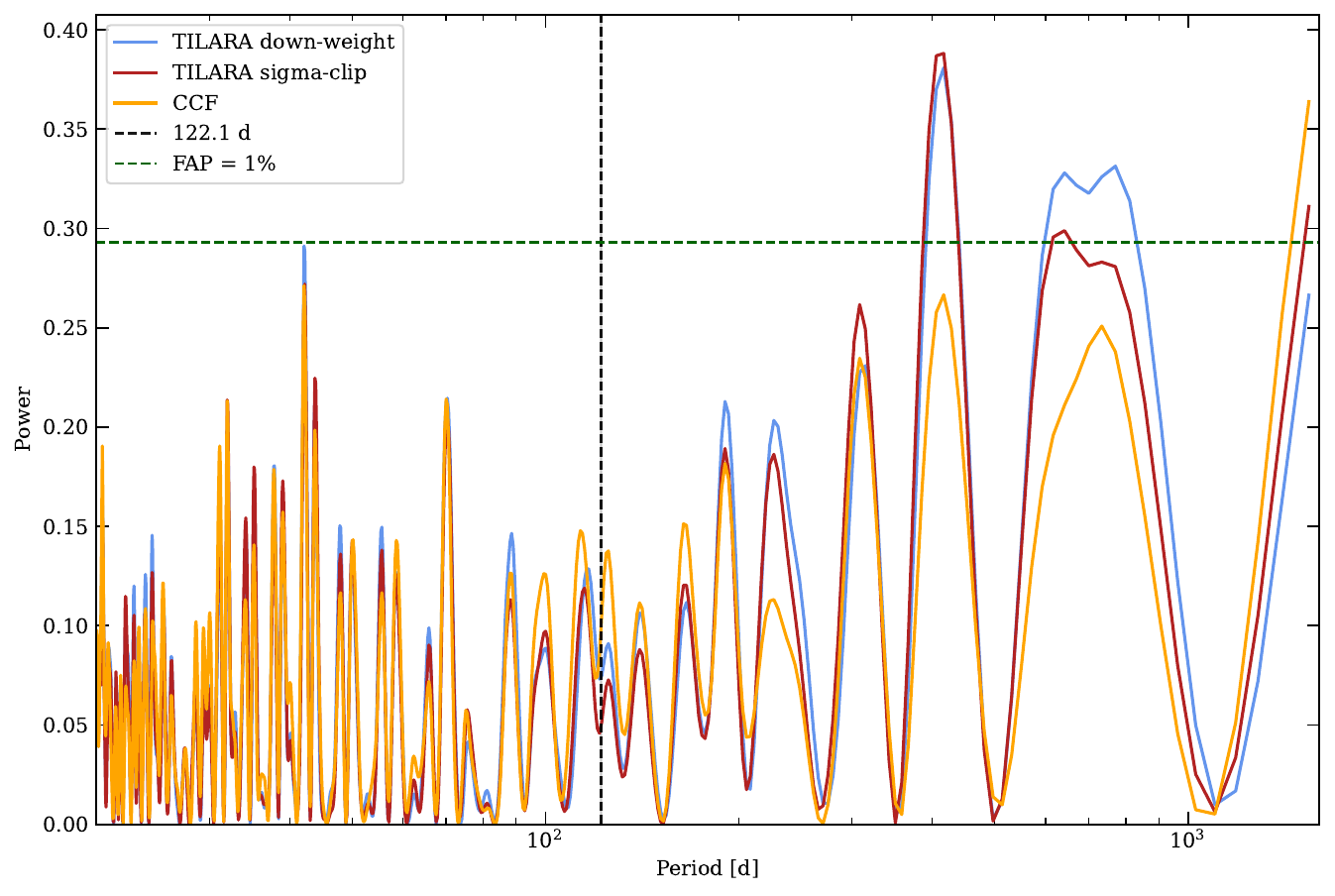}
   \caption{Generalized Lomb–Scargle periodograms of HD~102365 RVs. The orange curve corresponds to CCF-derived RVs, while the blue curve shows RVs obtained with \texttt{TILARA} using down-weighting and the red curve shows RVs obtained with \texttt{TILARA} using sigma-clipping. Horizontal dashed line indicate the 1$\%$ false alarm probability (FAP) for each dataset (for the three methods they overlap). The vertical dashed line and shaded region mark the expected orbital period of HD~102365 b (122.1 days ± 0.3 days). No significant peak is observed at the planet’s period in either dataset.}
              \label{Periodogram}%
    \end{figure*}
We computed a generalized Lomb–Scargle periodograms \citep{Zechmeister2009} for both the CCF-derived RVs and the RVs extracted with \texttt{TILARA}. As mentioned in Sec. \ref{HD~102365}, the time-series spans from the 5th of January 2019 to the 26th of March 2023, with a median S/N of 267, providing sufficient coverage to probe the orbital period of the previously reported Neptune-mass planet, HD~102365 b.

The periodograms were calculated on a dense frequency grid, accounting for the measurement error. The resulting power spectra are shown in Fig. \ref{Periodogram}. Horizontal dashed lines indicate the 1$\%$ false alarm probability (FAP) levels for each dataset. The expected orbital period of HD~102365 b (122.1 days) is marked with a vertical dashed line and a shaded ±0.3 day region.

Neither the \texttt{TILARA}-derived nor the CCF-derived periodogram show a significant signal at the expected 122.1-day period above the 1$\%$ FAP, indicating a non-detection of HD~102365 b in this dataset which indicates that no significant periodic signal is detected at this period in the present dataset. However, the absence of a significant peak in the periodogram alone is not sufficient to rule out the presence of a planetary signal. Moreover, the work of \cite{Figueira2025} shows that the ESPRESSO data on their own do not provide conclusive evidence to support the existence of planets around HD~102365. Since the \texttt{TILARA} RVs are consistent with the CCF and template-matching RVs analyzed in \cite{Figueira2025}, they are expected to lead to the same conclusions when subjected to a similarly detailed analysis.
Our analysis, together with the work done in \cite{Figueira2025}, is therefore consistent with the interpretation that, within the precision limits of ESPRESSO, the signal previously attributed to HD~102365 b is dubious.

The periodogram of the \texttt{TILARA} RVs shows that the peaks are more significant compared to the CCF RVs. This behavior is expected, as the line-by-line approach preserves line-dependent RV signals that may be attenuated when averaging over many lines in the CCF method. These more prominent signals may be associated with stellar activity, wavelength-dependent effects, or other systematics, and should not be interpreted without further analysis.

\section{Summary and conclusions}\label{sec:conclusions}

We have presented \texttt{TILARA}, a template-independent, line-by-line RV extraction code designed to work in conjunction with \texttt{ARES}. The code is motivated by the upcoming PoET instrument \citep{Santos2025}, which will enable disk-resolved solar observations, offering unprecedented opportunities to study small-scale phenomena such as granulation, supergranulation, and sunspots in the context of RV measurements.  

A key motivation behind \texttt{TILARA} is the need to measure \emph{true} line-by-line RVs, rather than velocities associated with broad spectral segments defined between consecutive maxima, as done in many existing line-by-line approaches. Because \texttt{ARES} can resolve blended features, \texttt{TILARA} derives velocities from physically distinct absorption lines, enabling a direct link between RVs and the physical properties of each line (e.g. formation height, excitation potential, magnetic sensitivity). This opens the door to identifying which lines are more or less affected by stellar activity, a critical step toward activity-mitigated RV extraction.

\texttt{TILARA} incorporates (but is not limited to) a carefully constructed line list refined from ESPRESSO solar observations, and implements flexible tools for error estimation, outlier removal, and line weighting. Its modular design allows it to be adapted to different spectral types and observational datasets. The code follows the \cite{Bouchy2001} method for RV error calculations, and supports both classical sigma-clipping and down-weighting to mitigate the impact of spectral outliers.  

Applying \texttt{TILARA} to a four-year ESPRESSO dataset of the solar-type star HD~102365 demonstrated that the  \texttt{TILARA} approach can achieve a comparable RV scatter to the standard \texttt{CCF} method and other state-of-the-art pipelines, while obtaining slightly lower average error bars (see Table~\ref{tab:rv_stats}). These results confirm that precise and robust RV measurements can be obtained without relying on spectral templates, an important advantage when such templates are difficult or impossible to construct. \texttt{TILARA} combines the template-free advantage of the \texttt{CCF} extraction technique with the outlier-resistant advantage of the line-by-line extraction technique.  

Given its flexibility, \texttt{TILARA} could in theory be applied to a range of stellar targets and observational programs. In particular, its ability to down-weight or remove problematic lines without discarding entire spectra should make it suitable for stars with complex or variable activity signals which we intend to explore in upcoming studies. 

Future work will focus on integrating \texttt{TILARA} with PoET’s disk-resolved datasets to investigate how localized stellar surface features contribute to the RV variability of individual lines. This approach is particularly critical for high-spatial-resolution observations of phenomena such as granulation and supergranulation. In these cases, the stellar spectrum varies so significantly across the solar disk and over short timescales that a stable, representative reference template -- the fundamental requirement for cross-correlation and the other methods used for comparison in this work -- cannot be reliably constructed. Unlike these traditional methods, which assume a fixed spectral shape to measure Doppler shifts, \texttt{TILARA} operates independently of a flux template. This allows for the robust extraction of line-specific RV variations even in the presence of rapidly evolving surface phenomena, providing a pathway to disentangle stellar activity from planetary signals by studying the physical behavior of individual spectral features.

While \texttt{TILARA} demonstrates high precision for Sun-like stars, several considerations must be addressed when extending its application to other stellar types or datasets. Currently, the algorithm is optimized for quiet, slowly rotating G-type stars with well-defined, narrow absorption lines. For cooler stars, such as M-dwarfs, the high density of spectral lines leads to significant blending, which poses a challenge for the Gaussian decomposition performed by \texttt{ARES}. In such regimes, the use of alternative line models, restricted wavelength ranges, or higher spectral resolution may be required to maintain RV accuracy.

Furthermore, the construction of the reference line list is a critical step that is not yet universal. The stability of a list depends heavily on the star’s spectral type, metallicity, age, and rotation rate. Consequently, we recommend the use of star-specific line lists rather than transferring lists across different stellar populations. Regarding the data requirements, while a small number of high S/N spectra may be sufficient to establish a reference for quiet stars, targets with significant intrinsic variability or unknown planetary signals may necessitate larger datasets and iterative refinement to ensure that the reference line centers are not biased by transient phenomena.

It is also important to contextualize \texttt{TILARA}’s role relative to state-of-the-art template-matching methods. When a high-quality, stable stellar template can be constructed, template-matching generally extracts more RV information and achieves lower residuals than Gaussian-based approaches. \texttt{TILARA} is not designed to replace these methods in those specific cases. Instead, it provides a complementary, line-by-line framework that preserves the individual information of each spectral feature. This flexibility is essential for studying activity-related phenomena -- such as the correlation between RV variations and line formation depths -- which are inherently difficult to isolate using global template-matching techniques that deliver a single, integrated RV value.

Overall, this work highlights the potential of the \texttt{TILARA} approach as a powerful complement to traditional CCF-based and template-matching RV extraction methods, particularly in the era of ultra-high precision spectroscopy.

The \texttt{TILARA} code, along with the solar line list and documentation, is publicly available on GitHub at \url{https://github.com/Carmensnm/TILARA}. Users are encouraged to adapt the modular framework for different instruments and stellar targets.

\begin{acknowledgements}
This work was funded by the European Union (ERC, FIERCE, 101052347). Views and opinions expressed are however those of the author(s) only and do not necessarily reflect those of the European Union or the European Research Council. Neither the European Union nor the granting authority can be held responsible for them. This work was also supported by FCT - Fundação para a Ciência e a Tecnologia through national funds by these grants: UIDB/04434/2020 DOI: 10.54499/UIDB/04434/2020, UIDP/04434/2020 DOI: 10.54499/UIDP/04434/2020, PTDC/FIS-AST/4862/2020, UID/04434/2025.

K.~A. acknowledges support from the Swiss National Science Foundation (SNSF) under the Postdoc Mobility grant P500PT\_230225.

V.~A. acknowledges support from FCT through a work contract funded by the FCT Scientific Employment Stimulus program (reference 2023.06055.CEECIND/CP2839/CT0005, DOI: 10.54499/2023.06055.CEECIND/CP2839/CT0005).
\end{acknowledgements}

%
%

\bibliographystyle{mnras}
\bibliography{refs.bib}
\begin{appendix} 

\clearpage
\section{Technical details of \texttt{linesearcher}} \label{app:line_selection}
Before running \texttt{linesearcher}, the S1D FITS files were pre-processed by interpolating the spectrum onto a regular wavelength grid with a constant step of 0.005~\AA. While modern ESPRESSO S1D files are stored in FITS tables that allow for slightly varying dispersion across the wavelength range, a uniform sampling is required in this step to ensure it is compatible with the necessary input in \texttt{ARES}. This interpolation step matches the nominal numerical resolution of the instrument while providing the regular grid necessary for the code to operate consistently across the entire spectral range. This uniform sampling is required to ensure accurate line detection.

The key parameters controlling the detection process in \texttt{linesearcher} are:  
\begin{itemize} 
\item \textbf{\texttt{lambdai $\&$ lambdaf}}: define initial and final wavelength values of the spectral interval over which the search is carried out.  
\item \textbf{\texttt{smoothder}}: a smoothing factor applied when computing numerical derivatives, helping to suppress noise and improve the detection of absorption features.  
\item \textbf{\texttt{rejt}}: a continuum normalization parameter. During normalization, a second-order polynomial is iteratively fit to the spectrum, with only points above a threshold (set by \texttt{rejt}) retained for the fit. This value could be adapted to the S/N of the spectrum.  
\item \textbf{\texttt{lineresol}}: the minimum allowed separation between two detected lines. If two features are found closer than this limit, they are treated as a single blended line.  
\end{itemize}

The output of \texttt{linesearcher} is a text file with two columns: the first column lists the estimated central wavelengths of the detected lines, and the second gives the corresponding depths.  

This preliminary list is then used as input for \texttt{ARES}, which refines the line positions by fitting Gaussians to each candidate feature. \texttt{ARES} automatically measures EWs and their errors, line depths, and FWHM \citep{Sousa2007}. For each candidate line, the program selects a local spectral window, determines the continuum level, and computes the best Gaussian fit. 

\section{Technical details of \texttt{ARES}} \label{app:ARES}

\texttt{ARES} requires as input a S1D spectrum, a list of approximate line centers to initialize the fits, and a set of user-defined parameters that control the fitting process.  

Most of the input parameters are shared with \texttt{linesearcher}, but \texttt{ARES} also includes additional options that are critical for refining the line fits:  
\begin{itemize}
    \item \textbf{\texttt{space}}: the wavelength interval (in \AA) around each line considered for the Gaussian fit. 
    \item \textbf{\texttt{miniline}}: the minimum line depth required for the line to be considered and fitted.  
    \item \textbf{\texttt{rvmask}}: the velocity shift (in km\,s$^{-1}$) applied to the line mask to align it with the observed spectrum.  
\end{itemize}

The standard output of \texttt{ARES} is a text file with the extension \texttt{.ares}, containing nine columns with detailed information for each line:  
\begin{enumerate}
    \item Wavelength of the line.  
    \item Number of components fitted to the line.  
    \item Depth of the line.  
    \item FWHM of the line.  
    \item EW of the line.  
    \item Uncertainty of the EW.  
    \item Depth of the fitted Gaussian.  
    \item Width (standard deviation) of the fitted Gaussian.  
    \item Central wavelength of the fitted Gaussian.  
\end{enumerate}

Here, the \textit{wavelength of the line} refers to the reference wavelength determined in Step 1 from the homogenized line list and it is used as input for\texttt{ARES} to define the initial line position and fitting window. In contrast, the “central wavelength of the fitted Gaussian” corresponds to the line center returned by\texttt{ARES} after fitting a Gaussian profile to the observed absorption feature.

All measured parameters -- namely the line center, EW, depth, and FWHM -- are derived from the best-fit Gaussian profile. \texttt{ARES} identifies and resolves blended features by searching for multiple local minima within the wavelength interval defined by the \texttt{space} parameter; when more than one feature is detected, a simultaneous multi-Gaussian fit is performed.
The primary limitation for \texttt{ARES} arises in cases of extreme blending, where multiple fitted components are centered at nearly the same wavelength\footnote{We refer the reader to the original\texttt{ARES} papers \citep{Sousa2007, Sousa2015} for the exhaustive details of the blend-detection algorithm.}. In such situations, although the total EW is generally robust, individual parameters such as the FWHM or depth may not accurately reflect the physical properties of a single spectral line. Nevertheless, the line-by-line nature of \texttt{TILARA} enables the identification and exclusion of these problematic features from the final radial-velocity computation based on fit quality criteria or physical consistency checks.

This information forms the basis for the line-by-line RV calculations described in Sect.~\ref{RVanderrors}.

\section{Rejt selection}\label{rejt}
As mentioned in Appendix \ref{app:line_selection}, the rejt is a continuum normalization parameter that defines the threshold above which points are retained during the polynomial fitting of the spectrum. \texttt{ARES} performs a simple local continuum normalization as described in Section 3.1 of \cite{Sousa2007}, and since no global spectral normalization is applied, no spectral lines are excluded during this process; as expected, the choice of normalization affects the measured line depths and therefore the optimal selection of the rejt parameter is crucial.

The objective of this Section was to test different methods to select the optimal value of the \texttt{rejt} parameter.

Before obtaining the final \texttt{ARES} results of the list of reference lines and performing the statistical analysis mentioned in the previous subsection, we tested five different approaches for determining the optimal values of the \texttt{rejt} parameter in \texttt{ARES}, as well as to study the impact of the different options. The first three approaches used fixed values based on the mean S/N of the observations (S/N~$\approx$~260), as follows:
\begin{enumerate}
    \item Mean S/N value ($\approx$~260).
    \item Half the mean S/N ($\approx$~130).
    \item One-quarter of the mean S/N ($\approx$~65).
\end{enumerate}

\cite{Sousa2015} also describe two alternative strategies:
\begin{enumerate}
    \setcounter{enumi}{3}
    \item Estimate \texttt{rejt} from line-free regions of the spectrum. For our case, we adopted the wavelength intervals recommended by \cite{Sousa2007}: 5764--5766~\AA, 6047--6052~\AA, and 6068--6076~\AA.
    \item Make the \texttt{rejt} wavelength-dependent. This was implemented by measuring the S/N at the central wavelength of each order in each observation. Since ESPRESSO spectra consist of 85 echelle orders, we obtained 85 wavelength–S/N pairs per observation.
\end{enumerate}

For the same observation, we applied the 5 different \texttt{rejt} options, and obtained:
\begin{itemize}
    \item For an S/N $\approx$~260, 4310 lines
    \item For an S/N $\approx$~130, 4250 lines

    \item For an S/N $\approx$~65, 4207 lines

    \item For an S/N chosen by ranges, 4395 lines
    \item For an S/N varying with wavelength, 4290 lines
\end{itemize}
Interestingly, the variable method based on line-free regions of the spectrum yielded more lines than the rest of the methods. Also, the comparison of the three methods where the S/N was fixed, showed that higher S/N values generally resulted in a greater number of detected lines.  The large number of line counts at higher S/N values are consistent with the fact for a given spectrum, fixing the \texttt{rejt} to larger values would make the continuum placement higher as well, allowing more features to be interpreted as lines. For the rest of the paper, we adopted the fixed value corresponding to S/N~$\approx$~260, meaning that we keep 4310 lines. For comparison purposes, the \texttt{CCF} mask used in the official ESPRESSO pipeline uses 5484 lines.

\section{RV error methods} \label{appendixA}
Given the variety of methods available in the literature to compute RV errors, we explored four different approaches in this work.

Methods 1 and 2 take into account the idea that the precision of the RV will depend on the shape of the line as well as on the S/N. This precision is based on: the depth of the line, the S/N in the continuum and the width of the line. If we assume that the spectral lines will have a Gaussian shape, the error of the RV determined from a spectral line can be approximated as:
\begin{equation}
    \sigma_{RV} \sim \frac{\sqrt{FWHM}}{C \cdot S/N}
\end{equation}
where FWHM is the full width at half maximum per line, C is the contrast with the continuum (the depth divided by the continuum) and the S/N is the signal to noise ratio in the continuum, as depicted in \cite{Lovis2010}. This makes sense with the fact that the narrower and the deeper the line is, the smaller the error will be.  

Both Method~1 and Method~2 rely on this same expression for $\sigma_{RV}$; the only difference lies in how the S/N in the continuum is estimated. Method 1 calculates the S/N using interpolation. For this we used the values we calculated for the S/N when computing the \texttt{rejt}, and then we interpolated them to the nearby wavelengths. Method 2 approximates the S/N on the continuum using a simple formula to compute it based on the depth and the S/N in the center of the line:
\begin{equation}
    S/N_{continuum} = \frac{S/N_{line}}{1 - \text{Depth}} 
\end{equation}
where
\begin{equation}
    S/N_{line} = \frac{\text{$F$}}{\text{$F_{error}$}}
\end{equation}
Here, $S/N_{line}$ represents the S/N at the line center, computed as the ratio between the measured flux ($F$) and its associated uncertainty $F_{error}$ extracted from the observations.

Method 3 follows a completely different approach. It derives the Doppler formula, using the error of the center of the line produced by \texttt{ARES}, and since the reference wavelengths are taken as absolute (and using their errors would only produce a small offset), the formula for the error is:
\begin{equation}
    \sigma_{RV}=c \cdot \frac{\sigma_{\lambda_{\mathrm{obs}}}}{\lambda_{rest}}
\end{equation}
Lastly, Method 4 is explained in Sect.~\ref{RVanderrors}.

To select the best RV error calculation method, we computed the RVs of the example detailed in Sect.~\ref{HD~102365}, avoiding doing any sigma-clipping or down-weighting. We computed the average error bar of the RVs for the four different methods. The results obtained for the HD~102365 ESPRESSO 18 observations are summarized in Table~\ref{tab:rv_errors}. Based on these values, we decided
to discard Methods 2 and 3 since they seem to be underestimating the contribution of the errors. Lastly, we favored Method 4 (based on the \cite{Bouchy2001} formalism) over Method 1 due to the approximate nature of Method 1.

It is worth noting that for the purpose of these tests, the four error estimation methods were evaluated without applying the outlier rejection or down-weighting schemes described in Section \ref{weighting}. This approach was chosen to isolate the raw performance of each estimator and to ensure that the selected \cite{Bouchy2001} formalism provides the most statistically robust representation of the photon noise across the entire line list. Applying these filters prematurely would have masked the underlying behavior of the estimators by artificially removing the noisier lines that these formalisms are specifically designed to characterize.
\begin{table}[h!]
\centering
\caption{Average RV error bars for the ESPRESSO 18 HD~102365 (unbinned) results.}
\label{tab:rv_errors}
\begin{tabular}{lc}
\hline
\textbf{Method} & \textbf{Average Error [$m\,s^{-1}$]} \\
\hline
Method 1 & 0.17 \\
Method 2 & 0.02 \\
Method 3 & 0.01 \\
Method 4 & 0.13 \\
\texttt{CCF}      & 0.16 \\
\texttt{sBART} \citep{Silva2022} & 0.13 \\
\hline
\end{tabular}
\end{table}

\pagebreak

\section{Testing alternative down-weighting schemes} \label{down-weight-tests}

Before adopting the final probabilistic down-weighting scheme described in Sect.~\ref{Lorentz}, we tested several alternative functional forms to model the empirical distribution of line-to-line RV variability using the observations described in Sect.~\ref{HD~102365}. We first reproduced the method of \citet{Artigau2022}, fitting a Gaussian profile to the standardized RV values (mean-subtracted and $\sigma$-normalized). However, the Gaussian systematically underestimated the contribution of the heavy tails, indicating that it does not adequately represent the observed distribution (blue curve, top-left panel of Fig.~\ref{RV/RVerror}). A Lorentzian profile, which naturally features broader wings, provided a visibly better match to the data (red curve, same panel).

In a second test, instead of using the RV values (mean-subtracted and $\sigma$-normalized), we repeated the fitting procedure using the standard deviation of the RVs per line. In this case, we adopted truncated versions of both the Gaussian and Lorentzian functions, since the distributions start at zero and therefore have only one tail, forcing the fit to begin at the minimum of the distribution. The Gaussian again underestimated the tail contribution (top-right panel of Fig.~\ref{RV/RVerror}), whereas the truncated Lorentzian reproduced both the core and the wings more faithfully.

For each fit (Gaussian vs.\ Lorentzian, and RV values -- mean-subtracted and $\sigma$-normalized -- vs.\ RV standard deviation), we recomputed the final RV time-series and compared the resulting residuals for ESPRESSO~18 and ESPRESSO~19 (see bottom panels of Fig.~\ref{RV/RVerror}).
We also computed the standard deviation of the time-series obtained with both the Gaussian and the Lorentzian fit. When the down-weighting is based on the RV values (mean-subtracted and $\sigma$-normalized), both functions perform similarly, although the Lorentzian yields slightly lower scatter: 

\begin{itemize}
    \item \textbf{Gaussian}: 1.51\,m\,s$^{-1}$ for ESPRESSO~18 and 2.41\,m\,s$^{-1}$ for ESPRESSO~19.
    \item \textbf{Lorentzian}: 1.44\,m\,s$^{-1}$ for ESPRESSO~18 and 2.27\,m\,s$^{-1}$ for ESPRESSO~19.
\end{itemize}

The difference becomes more pronounced when the weighting is based on the RV standard deviation per line. In this case, the Lorentzian consistently provides the lowest residual dispersion:  

\begin{itemize}
    \item \textbf{Gaussian}: 1.47\,m\,s$^{-1}$ for ESPRESSO~18 and 2.17\,m\,s$^{-1}$ for ESPRESSO~19.
    \item \textbf{Lorentzian}: 1.20\,m\,s$^{-1}$ for ESPRESSO~18 and 2.10\,m\,s$^{-1}$ for ESPRESSO~19.
\end{itemize}
\label{app:periodogram_section}

Given its ability to better describe the heavy-tailed behavior of the empirical distributions and its consistently lower RV scatter, we adopted the truncated Lorentzian over the RV standard deviation per line as the default down-weighting function in \texttt{TILARA}.

\end{appendix}

\end{document}